\documentclass[floatfix, showpacs,showkeys,aps,prd,twocolumn]{revtex4-1}

\usepackage{lineno}
\usepackage{amsmath}
\usepackage{amssymb}

\usepackage{tikz}
\usepackage{lipsum,adjustbox}
\usepackage{graphicx}
\usepackage[caption = false]{subfig}
\usepackage{dcolumn}
\usepackage{bm}
\usepackage{gensymb}

\usepackage[inline]{enumitem}

\let\oldequation\equation
\let\oldendequation\endequation

\renewenvironment{equation}
  {\linenomathNonumbers\oldequation}
  {\oldendequation\endlinenomath}

\newcolumntype{L}[1]{>{\raggedright\arraybackslash}p{#1}}
\newcolumntype{C}[1]{>{\centering\arraybackslash}p{#1}}
\newcolumntype{R}[1]{>{\raggedleft\arraybackslash}p{#1}}

\usepackage{hyperref}
\hypersetup{
 pdftitle={},
 pdfauthor={},
 pdfsubject={},
 pdfkeywords={},
 pdfstartview={},
 bookmarksopen=true, breaklinks=true, debug=true,
 colorlinks=true, linkcolor=blue, citecolor=red, urlcolor=blue,
 hyperfigures=true
}

\begin{document}

\title{\bf \boldmath Observation of $\psi(3686)\to 3\phi$}

\author{
\begin{small}
\begin{center}
M.~Ablikim$^{1}$, M.~N.~Achasov$^{4,c}$, P.~Adlarson$^{75}$, O.~Afedulidis$^{3}$, X.~C.~Ai$^{80}$, R.~Aliberti$^{35}$, A.~Amoroso$^{74A,74C}$, Q.~An$^{71,58,a}$, Y.~Bai$^{57}$, O.~Bakina$^{36}$
% ...
\\
\vspace{0.2cm}
(BESIII Collaboration)\\
\vspace{0.2cm} {\it
$^{1}$ Institute of High Energy Physics, Beijing 100049, People's Republic of China\\
% ...
$^{3}$ Bochum  Ruhr-University, D-44780 Bochum, Germany\\
$^{4}$ Budker Institute of Nuclear Physics SB RAS (BINP), Novosibirsk 630090, Russia\\
% ...
$^{35}$ Johannes Gutenberg University of Mainz, Johann-Joachim-Becher-Weg 45, D-55099 Mainz, Germany\\
$^{36}$ Joint Institute for Nuclear Research, 141980 Dubna, Moscow region, Russia\\
% ...
$^{57}$ Southeast University, Nanjing 211100, People's Republic of China\\
$^{58}$ State Key Laboratory of Particle Detection and Electronics, Beijing 100049, Hefei 230026, People's Republic of China\\
% ...
$^{71}$ University of Science and Technology of China, Hefei 230026, People's Republic of China\\
% ...
$^{74}$ University of Turin and INFN, (A)University of Turin, I-10125, Turin, Italy; (B)University of Eastern Piedmont, I-15121, Alessandria, Italy; (C)INFN, I-10125, Turin, Italy\\
$^{75}$ Uppsala University, Box 516, SE-75120 Uppsala, Sweden\\
% ...
$^{80}$ Zhengzhou University, Zhengzhou 450001, People's Republic of China\\
\vspace{0.2cm}
$^{a}$ Deceased\\
% ...
$^{c}$ Also at the Novosibirsk State University, Novosibirsk, 630090, Russia\\
% ...
}\end{center}
\vspace{0.4cm}
\end{small}}
\date{\today}

\begin{abstract}
Using $(2.712\pm0.014)\times 10^9$ $\psi(3686)$ events collected by the BESIII detector operating at the BEPCII collider, we report the first observation of $\psi(3686)\to 3\phi$ decay with a significance larger than 10$\sigma$. The branching fraction of this decay is determined to be $(1.46\pm0.05\pm0.17)\times10^{-5}$, where the first uncertainty is statistical and the second is systematic. No significant structure is observed in the $\phi\phi$ invariant mass spectra.
\end{abstract}

\maketitle

\section{INTRODUCTION}\label{sec:intro}

Charmonium resonances lie in between the perturbative and non-perturbative regimes of Quantum Chromodynamics (QCD)~\cite{bes3-white-paper, qcd1, qcd2}, which describes the strong interaction.
Below the open charm threshold, both $J/\psi$ and $\psi(3686)$ mainly decay into light hadrons through the annihilation of the $c\bar{c}$ pair into three gluons or one single virtual photon, with the decay width proportional to the modules of the charmonium wave function~\cite{aspect3}. QCD has been tested thoroughly at high energy region where the strong interaction coupling constant is small. However, in the low energy region, theoretical calculations based on first principles of QCD are still unreliable since the non-perturbative contribution is significant, and various effective field theories are introduced~\cite{qcdtheory1, qcdtheory2, qcdtheory3} to approximate these non-perturbative contributions. The study of charmonium decays can provide valuable insights to improve the understanding of the inner charmonium structure and test phenomenological mechanisms of non-perturbative QCD.

In recent years, significant progress has been made in experimental studies of multi-body $J/\psi$ and $\psi(3686)$ decays.
Previously, the $\psi(3686)\to PPP$, $\psi(3686)\to VPP$, and $\psi(3686)\to VVP$ decays have been extensively studied, as summarized in Ref.~\cite{bes3-white-paper}, where $P$ and $V$ denote pseudoscalar and vector mesons, respectively.
To date, no study of $\psi(3686)\to VVV$ has been reported.
In this paper, we present the first observation of the $\psi(3686)\to 3\phi$ decay. This analysis is based on $(2.712\pm0.014)\times 10^9$ $\psi(3686)$ events collected at the center-of-mass energy of 3.686 GeV by the BESIII detector in 2009, 2012 and 2021~\cite{psip-num-inc}.

\section{BESIII EXPERIMENT AND MONTE CARLO SIMULATION}
The BESIII detector~\cite{2009MAblikimDet} records symmetric $e^+e^-$ collisions provided by the BEPCII storage ring~\cite{bepcii} in the center-of-mass energy ($\sqrt{s}$) range from 2.0 to 4.95~GeV, with a peak luminosity~($\mathcal{L}$) of $1 \times 10^{33}\;\text{cm}^{-2}\text{s}^{-1}$ achieved at $\sqrt{s} = 3.77\;\text{GeV}$. BESIII has collected large data samples in this energy region~\cite{bes3-white-paper,EcmsMea,EventFilter}. The cylindrical core of the BESIII detector covers 93\% of the full solid angle and consists of a helium-based multilayer drift chamber~(MDC), a plastic scintillator time-of-flight system~(TOF), and a CsI(Tl) electromagnetic calorimeter~(EMC), which are all enclosed in a superconducting solenoidal magnet providing a 1.0~T magnetic field. The solenoid is supported by an octagonal flux-return yoke with resistive plate counter muon identification modules interleaved with steel. 
The charged-particle momentum resolution at $1~{\rm GeV}/c$ is $0.5\%$, and the ${\rm d}E/{\rm d}x$ resolution is $6\%$ for electrons from Bhabha scattering. The EMC measures photon energies with a resolution of $2.5\%$ ($5\%$) at $1$~GeV in the barrel (end-cap) region. The time resolution in the TOF barrel region is 68~ps, while that in the end-cap region was 110~ps. The end-cap TOF system was upgraded in 2015 using multigap resistive plate chamber technology, providing a time resolution of 60~ps, which benefits about 85\% of the data used in this analysis~\cite{etof}.

Simulated data samples are produced with a {\sc geant4}-based~\cite{geant4} Monte Carlo (MC) package, which includes the geometric description of the BESIII detector and the detector response. The simulations model the beam energy spread and initial state radiation (ISR) in the $e^+e^-$ annihilations with the generator {\sc kkmc}~\cite{kkmc}. To estimate backgrounds, an inclusive MC sample is generated including the production of the $\psi(3686)$ resonance, the ISR production of the $J/\psi$, and the continuum processes incorporated in {\sc kkmc}~\cite{kkmc}. All particle decays are modelled with {\sc evtgen}~\cite{evtgen} using branching fractions either taken from the Particle Data Group~(PDG)~\cite{pdg2022}, when available, or otherwise estimated with {\sc lundcharm}~\cite{lundcharm}. Final state radiation from charged final state particles is incorporated using the {\sc photos} package~\cite{photos}. The detection efficiency of the $\psi(3686)\to 3\phi$ decay is determined using the signal MC samples containing $5\times10^5$ events, where the $\psi(3686)\to 3\phi$ and $\phi\to K^+K^-$ decays are generated with PHSP and VSS models, respectively. The PHSP model represents the generic phase space for n-body decays, averaging over the spins of initial and final state particles. The VSS model describes the decay of a vector particle $(\phi)$ into two scalar particles.

In addition, the data sample collected at the center-of-mass energy of 3.773 GeV with an integrated luminosity of 7.93~fb$^{-1}$~\cite{3773lumi} is used to estimate the contribution from continuum process.

\section{Event selection}

In this analysis, candidate events for $\psi(3686)\to 3\phi$ are selected by reconstructing three or two $\phi$ candidates. The two reconstruction methods are hereafter referred to as ``full reconstruction" for the three $\phi$ case and ``partial reconstruction" for the two $\phi$ case. A $\phi$ candidate is reconstructed by the decay $\phi \to K^+K^-$.

Each kaon candidate must satisfy $|\!\cos\theta| < 0.93$, where $\theta$ is the polar angle defined with respect to the $z$-axis, which is the symmetry axis of the MDC. Additionally, each kaon candidate must originate within $1$~cm ($10$~cm) of the interaction point in the plane transverse to the beam direction (in the beam direction).

Particle identification (PID) is performed on kaon candidates using the $\text{d}E/\text{d}x$ and TOF information. The charged kaons are identified by comparing the likelihoods for the kaon and pion hypotheses and requiring $\mathcal{L}(K)>\mathcal{L}(\pi)$.

\subsection{Full reconstruction sample}

A four-constraint (4C) kinematic fit, ensuring energy and momentum conservation, is performed under the hypothesis of $e^+e^-\to 3(K^+K^-)$ with at least six good kaon candidates. The helix parameters of charged tracks in the MC simulations are corrected to improve the $\chi^2$ distribution consistency between data and MC simulation using the method described in Ref.~\cite{helixmethod}. 
Events satisfying $\chi^2_{\rm{4C}}<50 $ are retained for further analysis. If there are multiple combinations in an event, the combination with the lowest $\chi^2_{\rm{4C}}$ is kept for further analysis.

The three $K^+K^-$ pairs result in six combinations to form the three different $\phi$ candidates. The best combination of three $\phi$ candidates is selected by minimizing
\begin{equation}
    \Delta = \sqrt{\sum_{i = a,b,c}(M^i_{K^+K^-}-m_{\phi})^2},
\end{equation}
where $m_\phi$ is the nominal $\phi$ mass~\cite{pdg2022}.
The three $\phi$ candidates are randomly labled by using the Knuth-Durstenfeld shuffle algorithm~\cite{durstenfield1964algorithm, knuth1969art}, since they are identical in the reconstruction procedure.

\subsection{Partial reconstruction sample}

To improve the detection efficiency, we employ the partial reconstruction strategy when six kaon candidates cannot be reconstructed. The reconstruction of the $\psi(3686)\to 3\phi$ decay is performed by selecting exactly five kaon candidates from six charged tracks in each event, with at least two kaons of each charge.
Events with less than five identified kaons are not used because of the very high combinatorial background.

The two $K^+K^-$ pairs along with one $K^\pm$ result in six combinations to form the two $\phi$ candidates. The best combination of two $\phi$ candidates is selected by minimizing
\begin{equation}
  \Delta = \sqrt{\sum_{i = a,b}(M^i_{K^+K^-}-m_{\phi})^2+ (M^{\rm rec}_{\phi}-m_{\phi})^2},
\end{equation}
where $M^{\rm rec}_{\phi}$ is defined as
\begin{equation}
  M^{\rm rec}_{\phi}=\sqrt{(\sqrt{s}-E_{\phi\phi})^2-p^2_{\phi\phi}},
\end{equation}
\noindent in which $E_{\phi\phi}$ and $p_{\phi\phi}$ are the energy and momentum of the two $\phi$ system, respectively. Additionally, to further improve the purity of the signal sample, the recoil mass of the $2(K^+K^-)K^\pm$ combination is required to be in the mass interval of $(0.4746,\ 0.5145)$ GeV/$c^2$. 
This range corresponds to about $\pm 3\sigma$ around the kaon mass, where $\sigma$ is the resolution
on the $2(K^+K^-)K^\pm$ recoil mass. Similar to the full reconstruction case, the two reconstructed $\phi$ candidates are randomly labled.

\subsection{Background analysis}

\begin{figure*}[htbp]\centering
  \includegraphics[keepaspectratio=true,width=0.495\textwidth,angle=0]{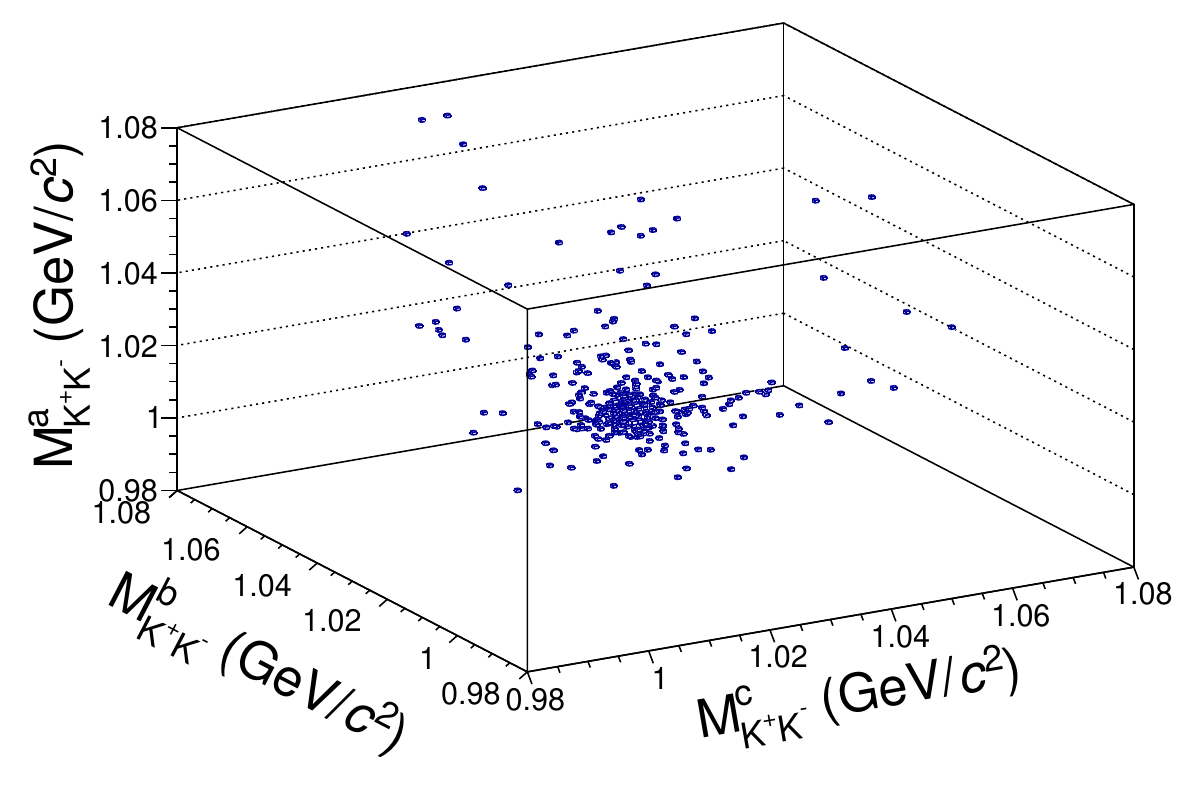}
  \includegraphics[keepaspectratio=true,width=0.495\textwidth,angle=0]{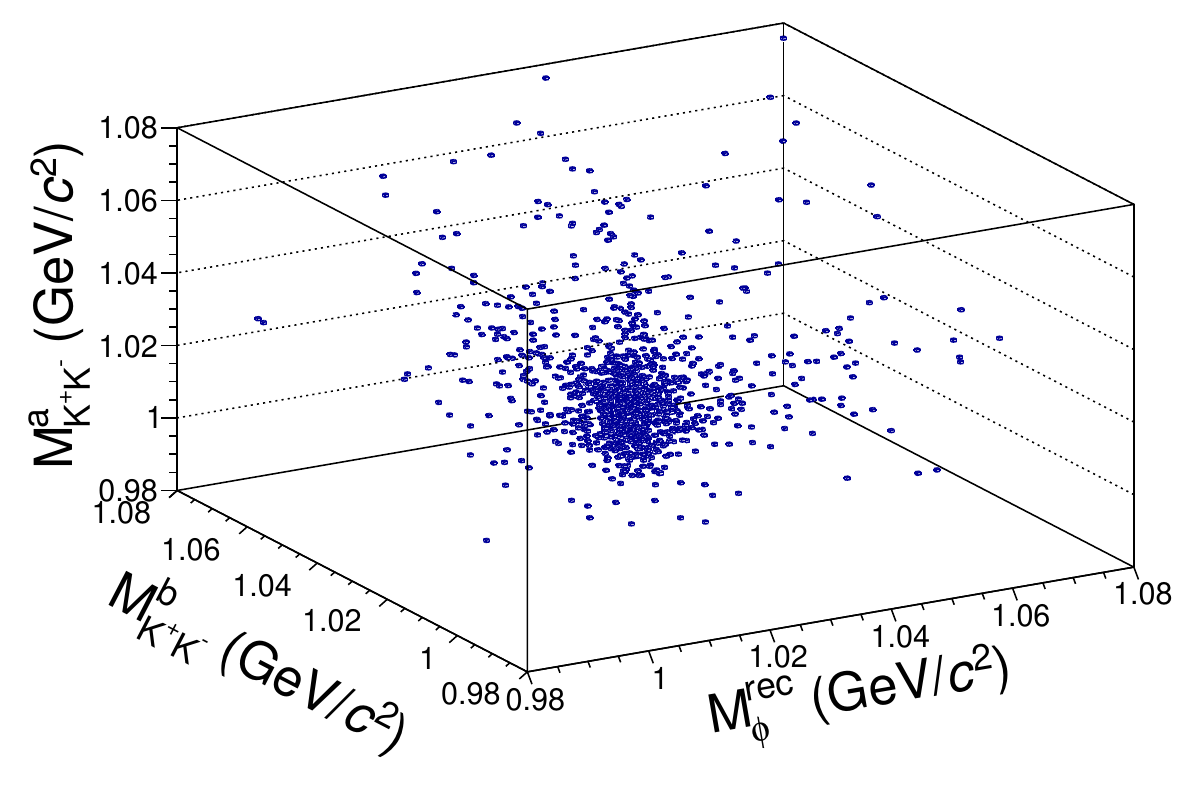}
  \caption{The $M^a_{K^+K^-}:M^b_{K^+K^-}:M^c_{K^+K^-}(M^{\rm rec}_{\phi})$ distributions of the (left) full and (right) partial reconstructed candidates for $\psi(3686)\to 3\phi$.}
  \label{fig:3d}
\end{figure*}

Potential background components are investigated by analyzing the inclusive MC sample of $\psi(3686)$ decays with the generic event type analysis tool, TopoAna~\cite{topo}. The study shows that only a very small background contribution survives the event selection. After imposing all selection criteria, the three dimensional (3D) distributions of the invariant masses of the three $\phi$ candidates in the data are shown in Fig.~\ref{fig:3d} for both full and partial reconstruction cases. In both cases, a distinct cluster around the $\phi$ mass is evident.

\section{Results}

\subsection{Fitting of data}

The signal yield of the $\psi(3686)\to 3\phi$ decay is determined through a simultaneous unbinned maximum likelihood fit to the 3D distribution of $M^a_{K^+K^-}:M^b_{K^+K^-}:M^c_{K^+K^-}$ for the full reconstruction case, and to the 3D distribution of $M^a_{K^+K^-}:M^b_{K^+K^-}:M^{\rm rec}_{\phi}$ for the partial reconstruction case. In the simultaneous fit, the branching fractions of $\psi(3686)\to 3\phi$ measured with the different reconstruction cases are constrained to be equal.
Events are divided into four cases based on the source of the $K^+K^-$ pairs:
the `Signal' describes candidates where all three pairs of $K^+K^-$ originate from $\phi$ mesons;
the `BKGI' denotes candidates where two pairs of $K^+K^-$ originate from $\phi$ mesons and one pair from combinatorial backgrounds;
the `BKGII' describes candidates where only one pair of $K^+K^-$ comes from the $\phi$ meson and the remaining two pairs are from combinatorial backgrounds;
the `BKGIII' encompasses candidates where all three pairs of  $K^+K^-$ come from combinatorial backgrounds, as well as the incorrectly reconstructed events with different final states.

So, the probability density functions (PDFs) of Signal, BKGI, BKGII, and BKGIII are constructed as follows:
\begin{itemize}
  \item Signal: ${\mathcal S}_x \times {\mathcal S}_y \times {\mathcal S}_z$,
  \item BKGI: $f_1\cdot{\mathcal S}_x \times {\mathcal S}_y  \times {\mathcal A}_z+{\mathcal S}_x \times {\mathcal A}_y  \times {\mathcal S}_z+{\mathcal A}_x \times {\mathcal S}_y  \times {\mathcal S}_z$,
  \item BKGII: ${\mathcal S}_x \times {\mathcal A}_y  \times {\mathcal A}_z+{\mathcal A}_x \times {\mathcal S}_y  \times {\mathcal A}_z+f_2\cdot{\mathcal A}_x \times {\mathcal A}_y  \times {\mathcal S}_z$,
  \item BKGIII: ${\mathcal A}_x \times {\mathcal A}_y  \times {\mathcal A}_z$.
\end{itemize}
Here, $x$, $y$, and $z$ correspond to the three dimensions of the 3D fit. The ${\mathcal S}_i$ are the signal shapes derived from signal MC simulations, while ${\mathcal A}_i$ are the reversed ARGUS functions~\cite{argus} that characterize the combinatorial background shape in the $K^+K^-$ invariant mass spectrum, where $i$ denotes the different dimensions. The parameters $f_1$ and $f_2$ describe the distinct PDFs resulting from the non-identical $\phi$ reconstructions in the partial reconstruction sample. This disparity is attributed to the resolution differences between the full and partial reconstructions. While $f_1$ and $f_2$ are fixed to 1 for the full reconstruction sample, they are treated as free fit parameters for the partial reconstruction sample to incorporate the resolution variation.

For the full reconstruction case, the ${\mathcal S}_i$ are identical and determined from the $\phi$ candidates of signal MC. Additionally, ${\mathcal A}_i$ are identical reversed ARGUS functions with starting points fixed at the $K^+K^-$ mass threshold.

For the partial reconstruction case, we specify the $z$ as the recoil dimension. ${\mathcal S}_x$ and ${\mathcal S}_y$ are identical PDFs derived from the two fully reconstructed $\phi$ candidates and ${\mathcal S}_z$ is derived from the line shape of $M^{\rm rec}_{\phi}$ of signal MC. ${\mathcal A}_x$ and ${\mathcal A}_y$ share the same parameters, while the parameters of ${\mathcal A}_z$ are independently determined. The starting points of all ${\mathcal A}_i$ are fixed  at the $K^+K^-$ mass threshold. 

\subsection{Detection efficiency}

\begin{figure*}[htbp]\centering
  \includegraphics[keepaspectratio=true,width=0.495\textwidth,angle=0]{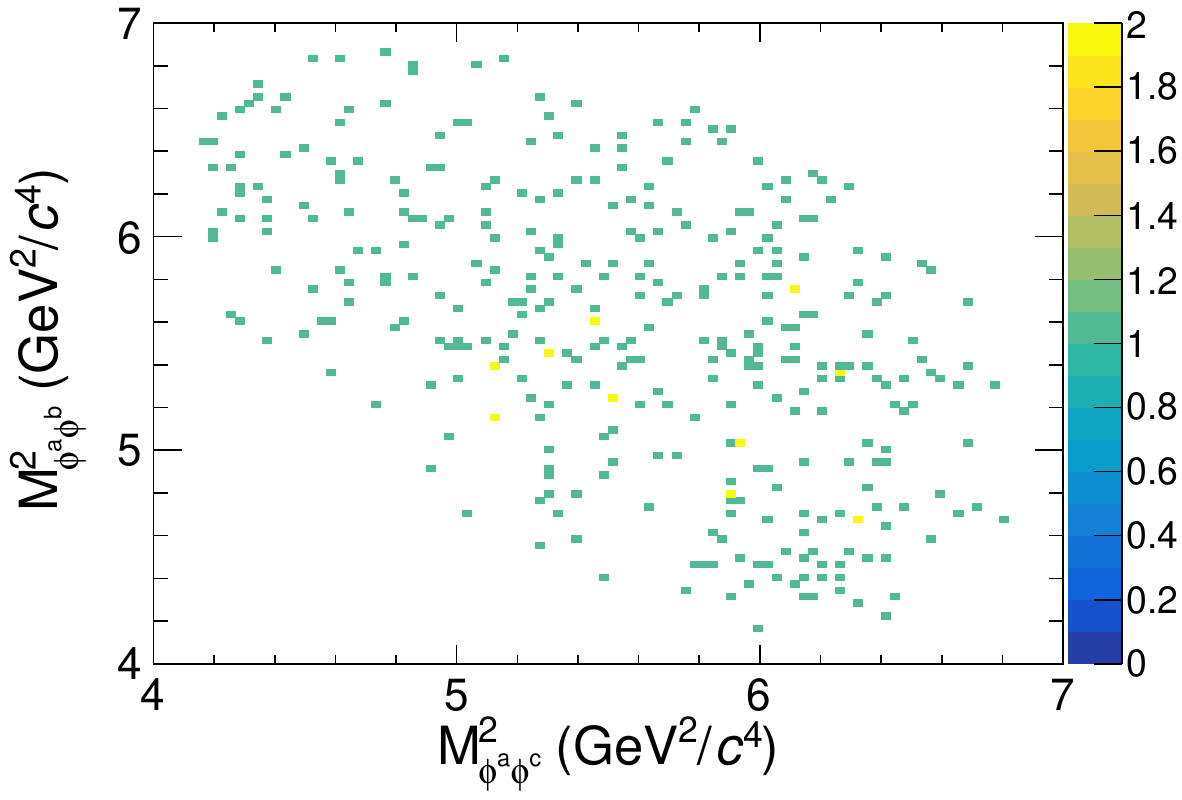}
  \includegraphics[keepaspectratio=true,width=0.495\textwidth,angle=0]{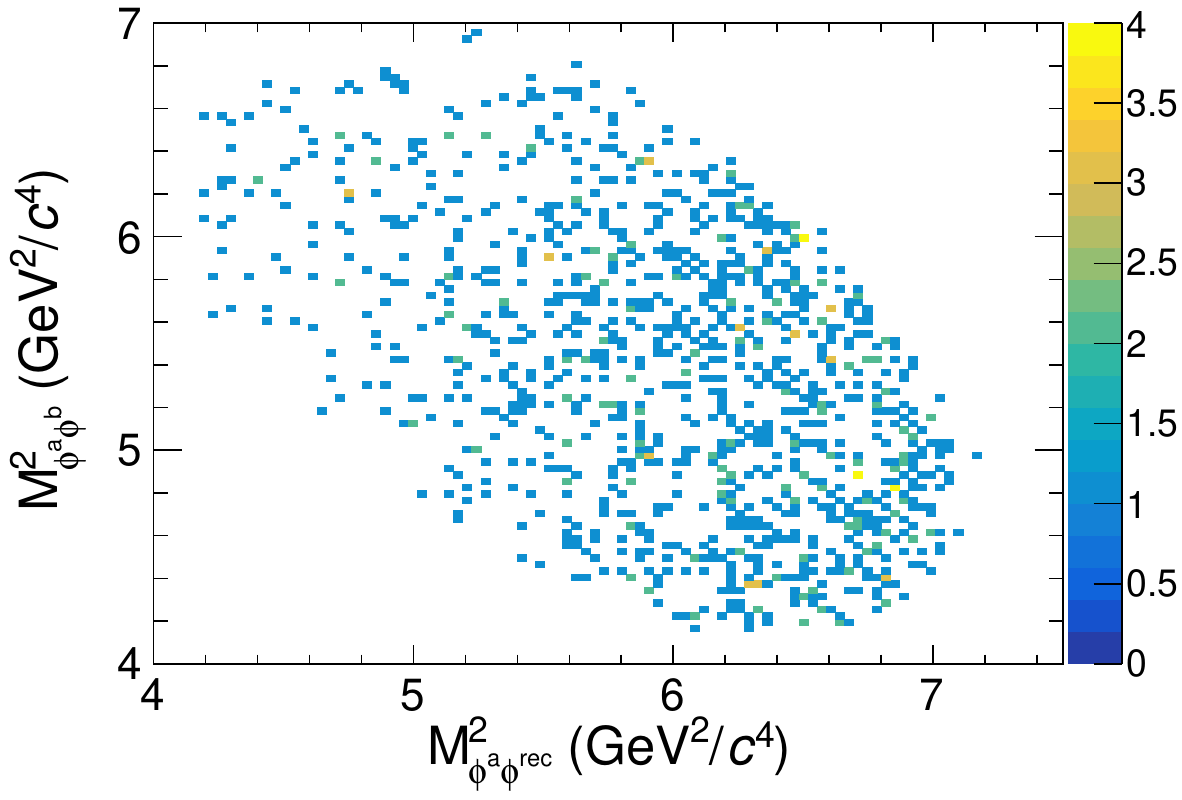}
  \includegraphics[keepaspectratio=true,width=0.495\textwidth,angle=0]{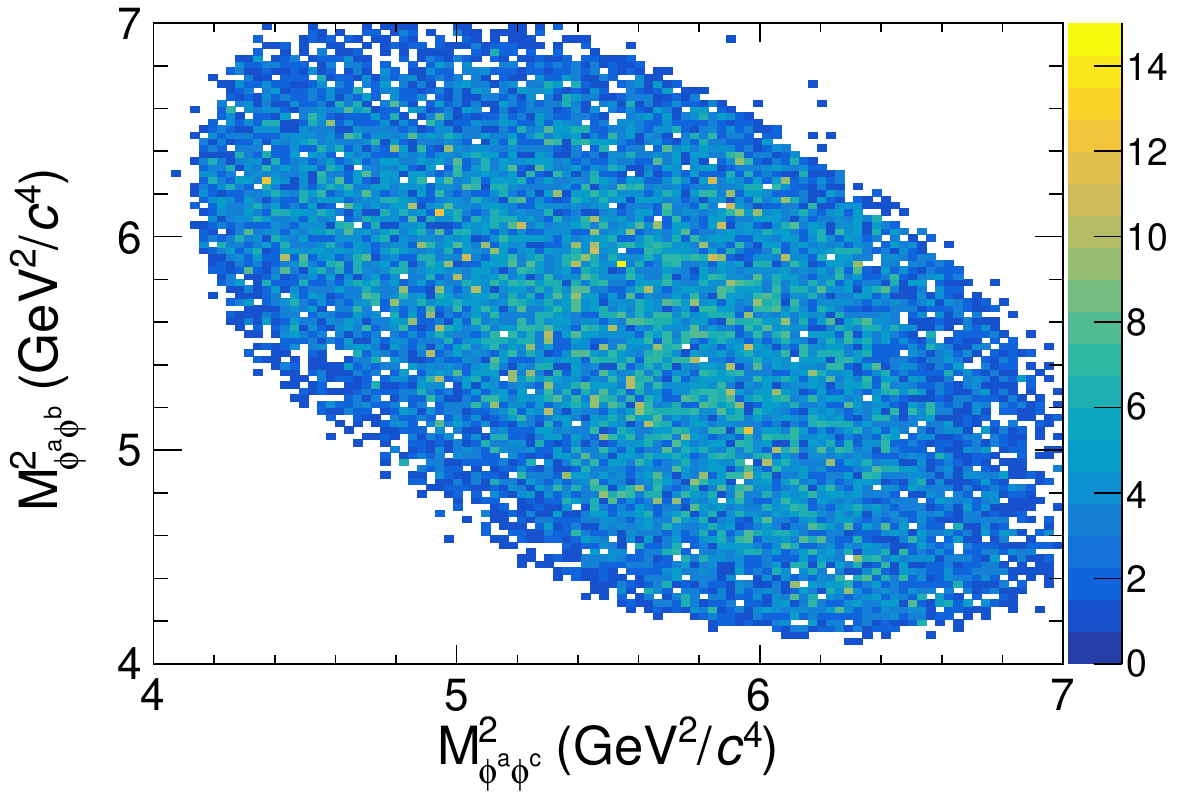}
  \includegraphics[keepaspectratio=true,width=0.495\textwidth,angle=0]{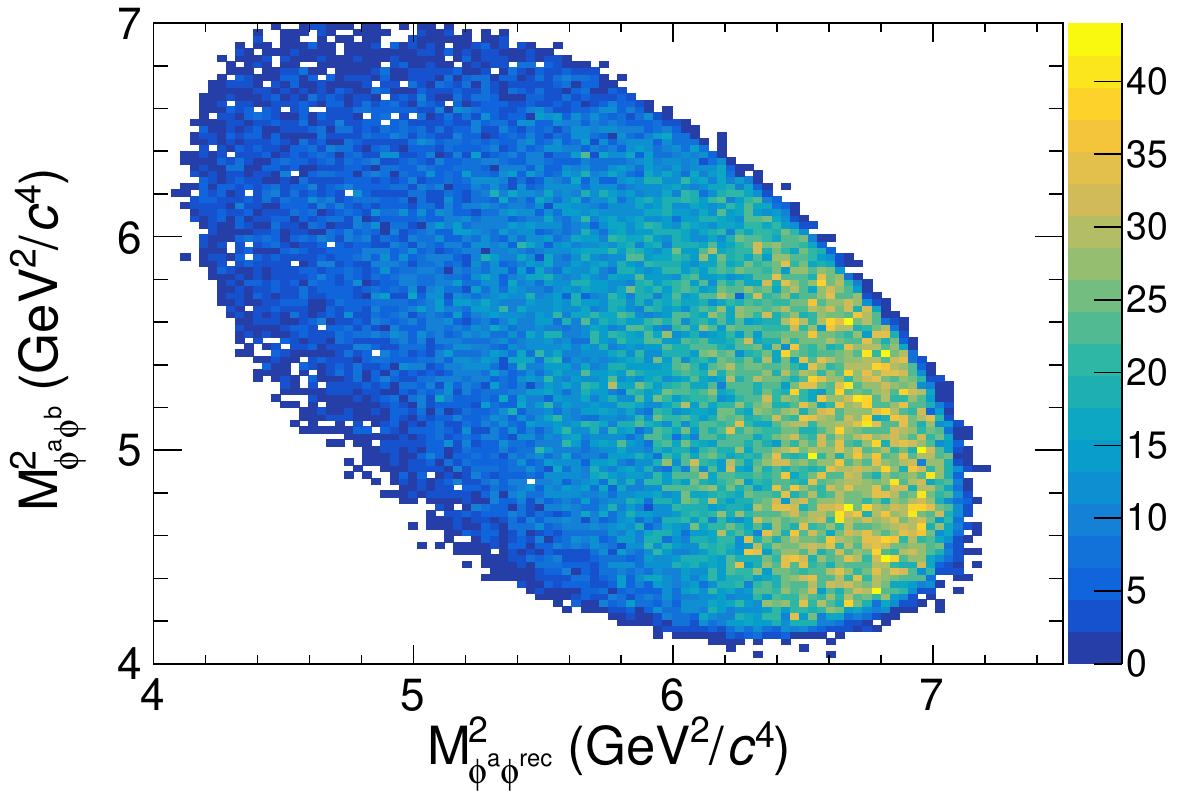}
  \includegraphics[keepaspectratio=true,width=0.495\textwidth,angle=0]{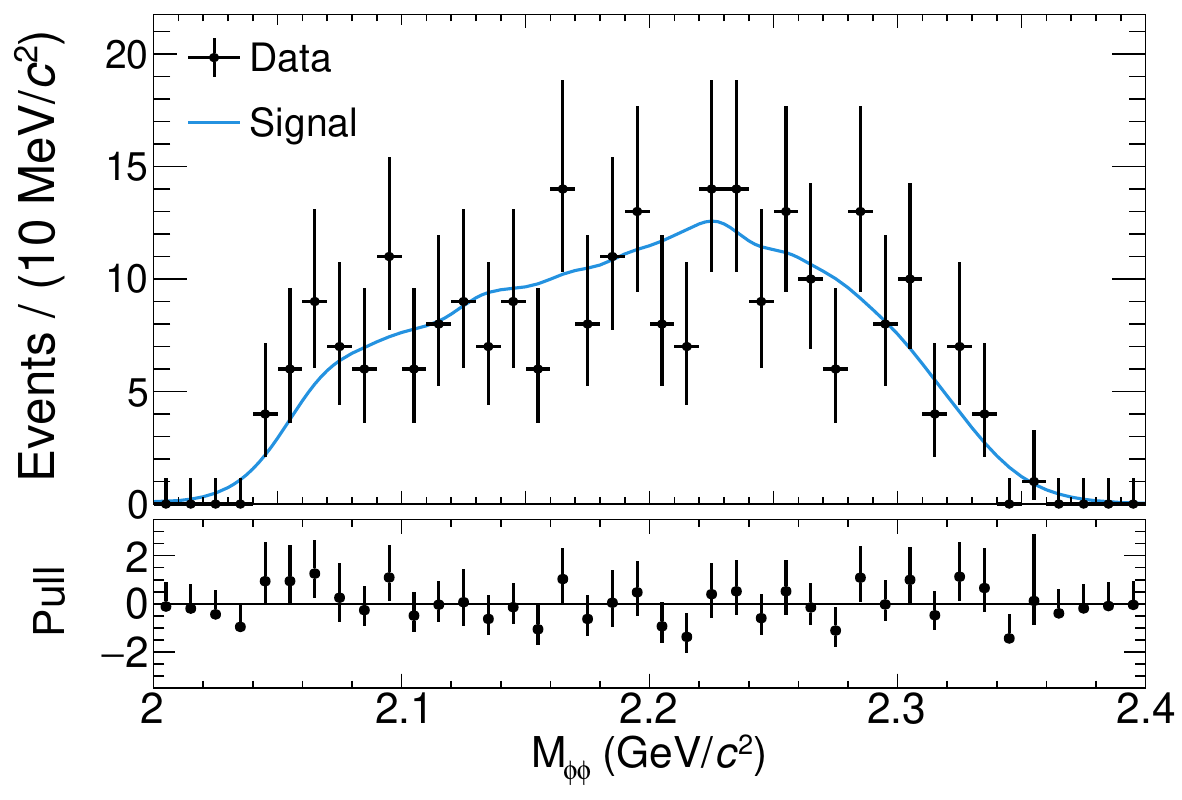}
  \includegraphics[keepaspectratio=true,width=0.495\textwidth,angle=0]{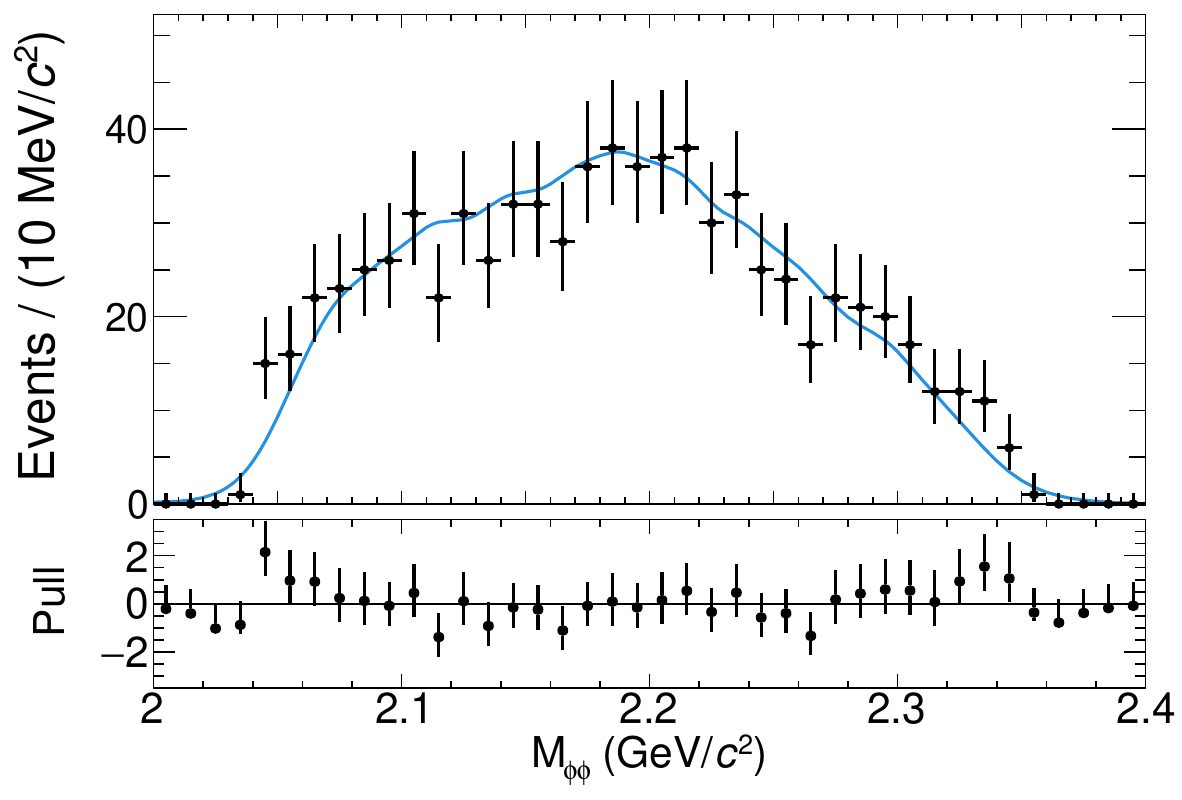}

  \caption{Dalitz plots of $ M^2_{\phi^a \phi^b}$ vs $ M^2_{\phi^a \phi^c}$ of the (left) full and (right) partial reconstructed candidates. The first row indicates data and the second row indicates signal MC events. In the third row, the distributions of the invariant masses of the two lowest momentum $\phi$ candidates are shown. The black points with error bars are data. The blue curves represent the signal component of the fit.}
  \label{fig:dalitz}
\end{figure*}

\begin{figure*}[htbp]\centering
  \includegraphics[keepaspectratio=true,width=0.325\textwidth,angle=0]{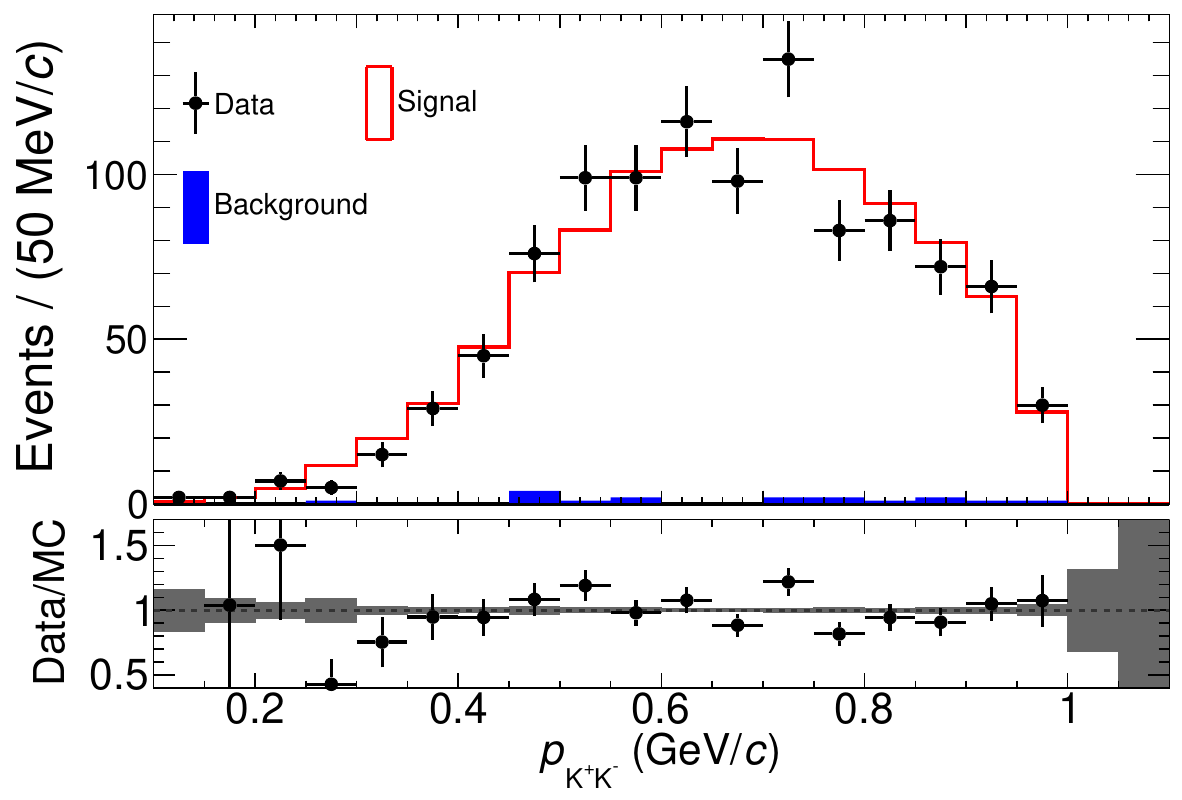}
  \includegraphics[keepaspectratio=true,width=0.325\textwidth,angle=0]{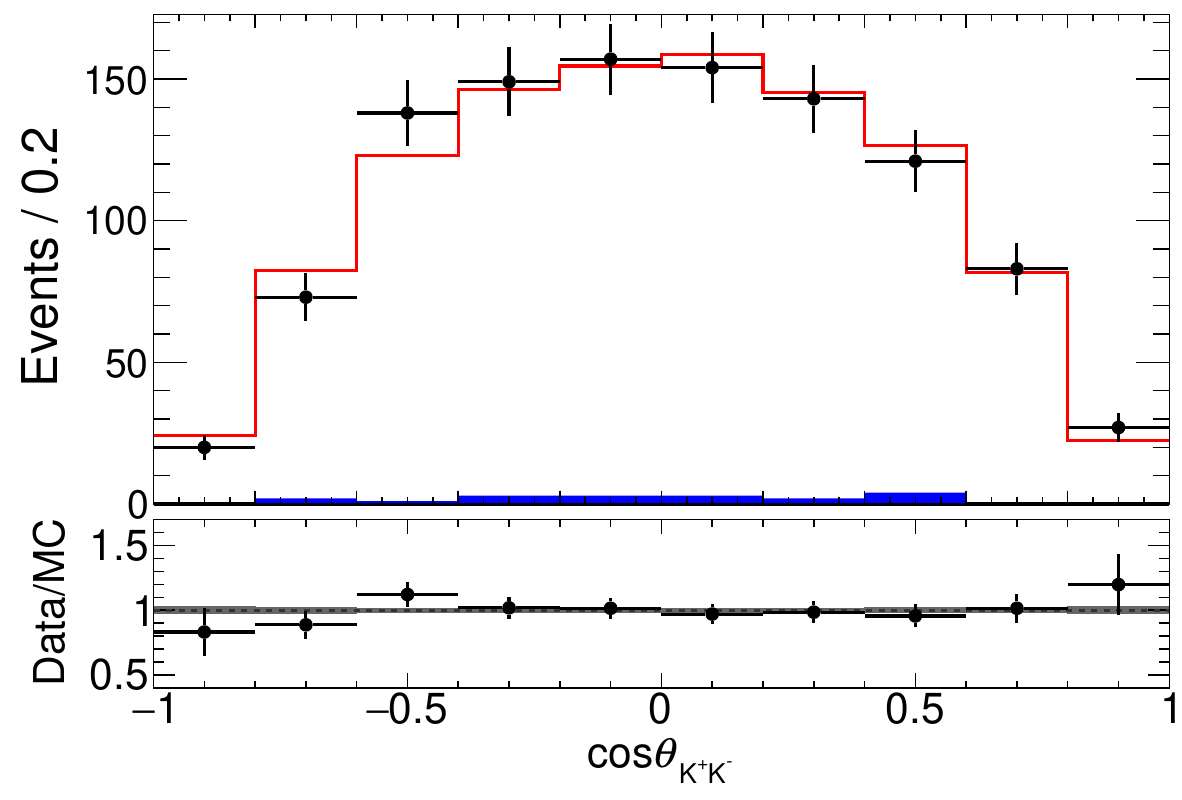}
  \includegraphics[keepaspectratio=true,width=0.325\textwidth,angle=0]{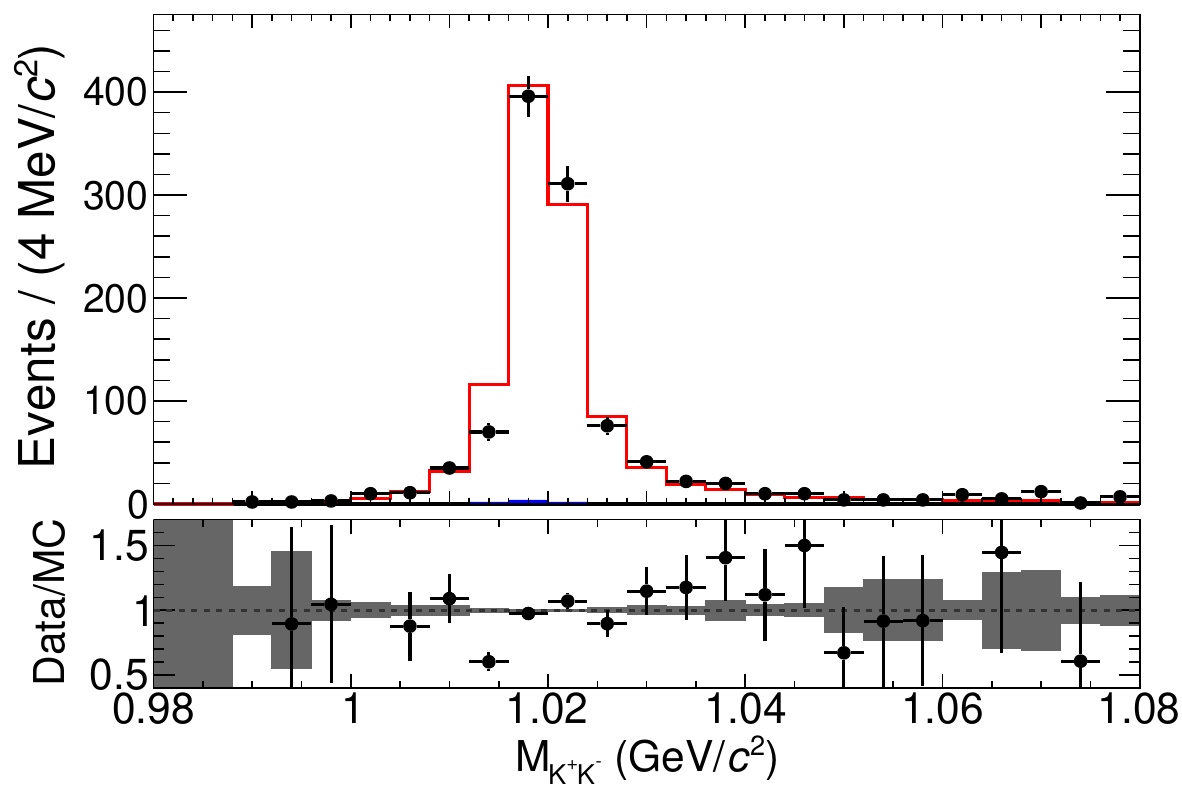}
  \caption{Distributions of the momentum, cosines of polar angle and mass spectra of the $\phi$ candidates for full reconstructed candidates, all reconstructed $\phi$ candidates are filled in the same histogram. The black points with error bars are data, the solid red lines show the signal MC simulation which is scaled to the total number of events of data, and the blue solid-filled histograms are the background contribution of the inclusice MC sample. The bottom panels show the data and MC comparison, where the error bands indicate the MC statistical uncertainty only.}
  \label{fig:compare6}
\end{figure*}

\begin{figure*}[htbp]\centering
  \includegraphics[keepaspectratio=true,width=0.495\textwidth,angle=0]{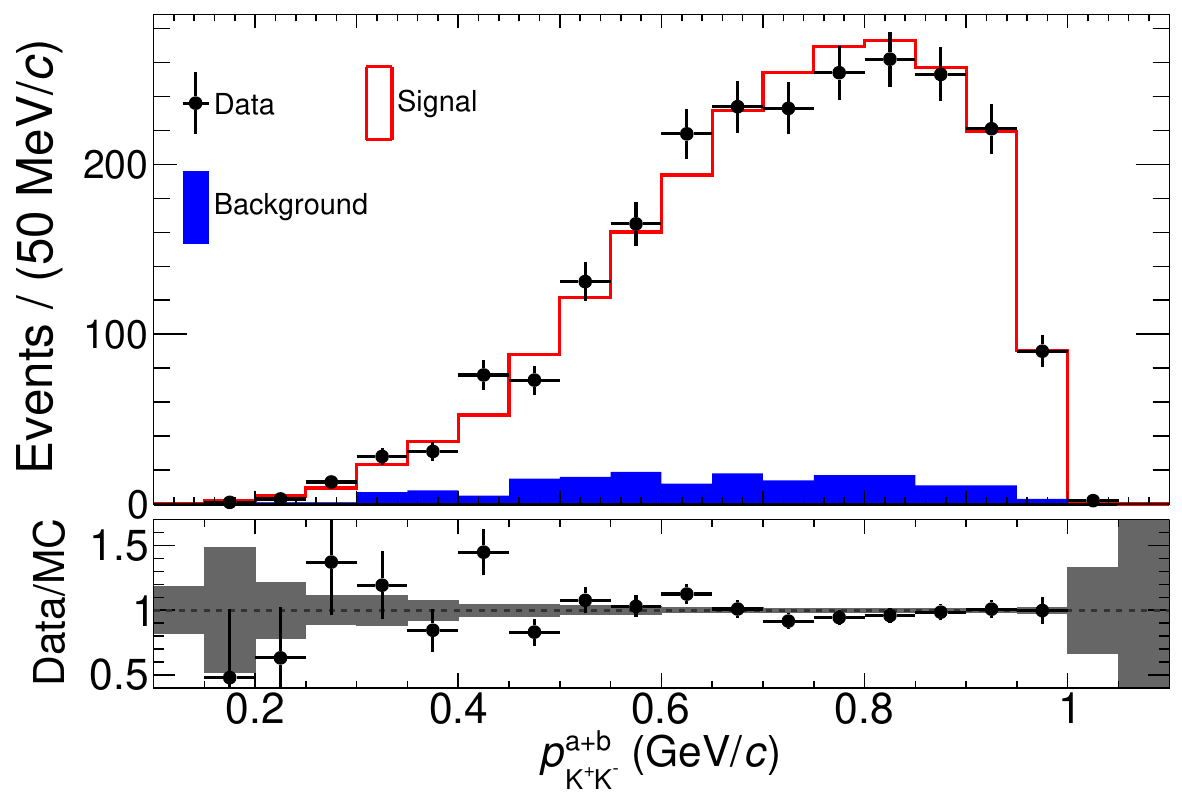}
  \includegraphics[keepaspectratio=true,width=0.495\textwidth,angle=0]{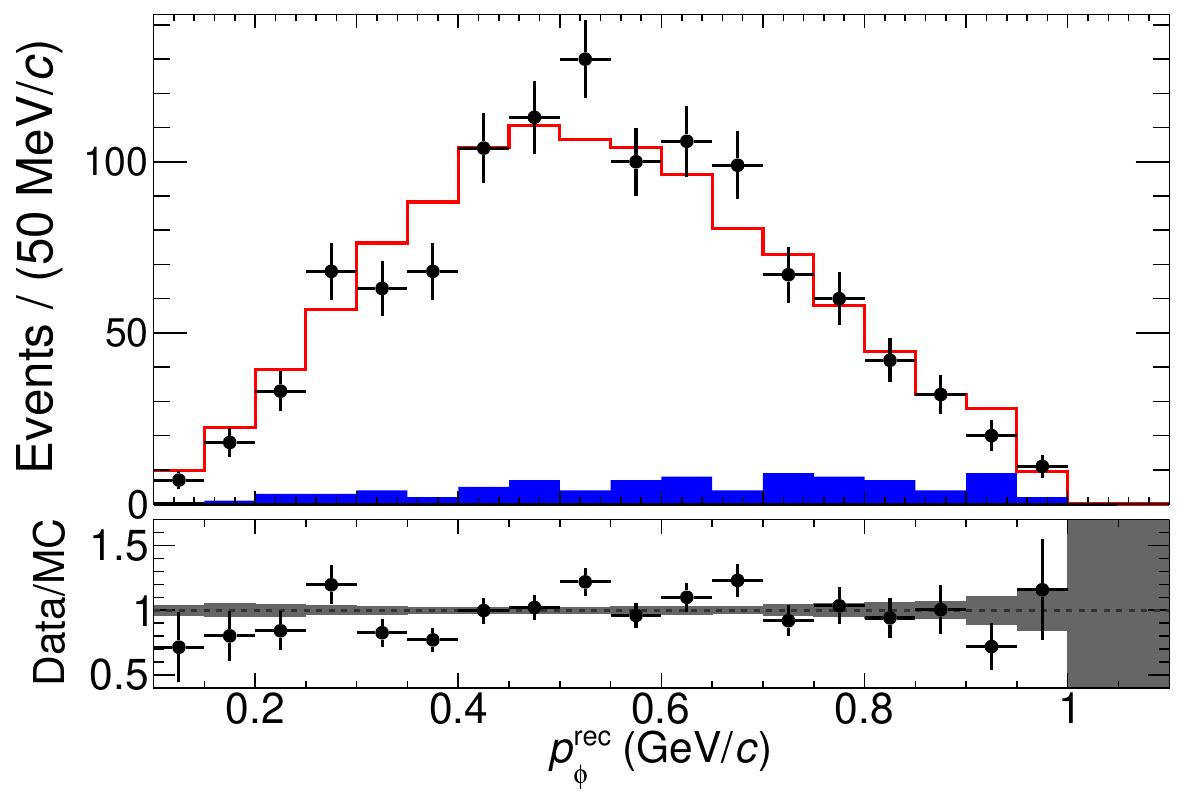}
  \includegraphics[keepaspectratio=true,width=0.495\textwidth,angle=0]{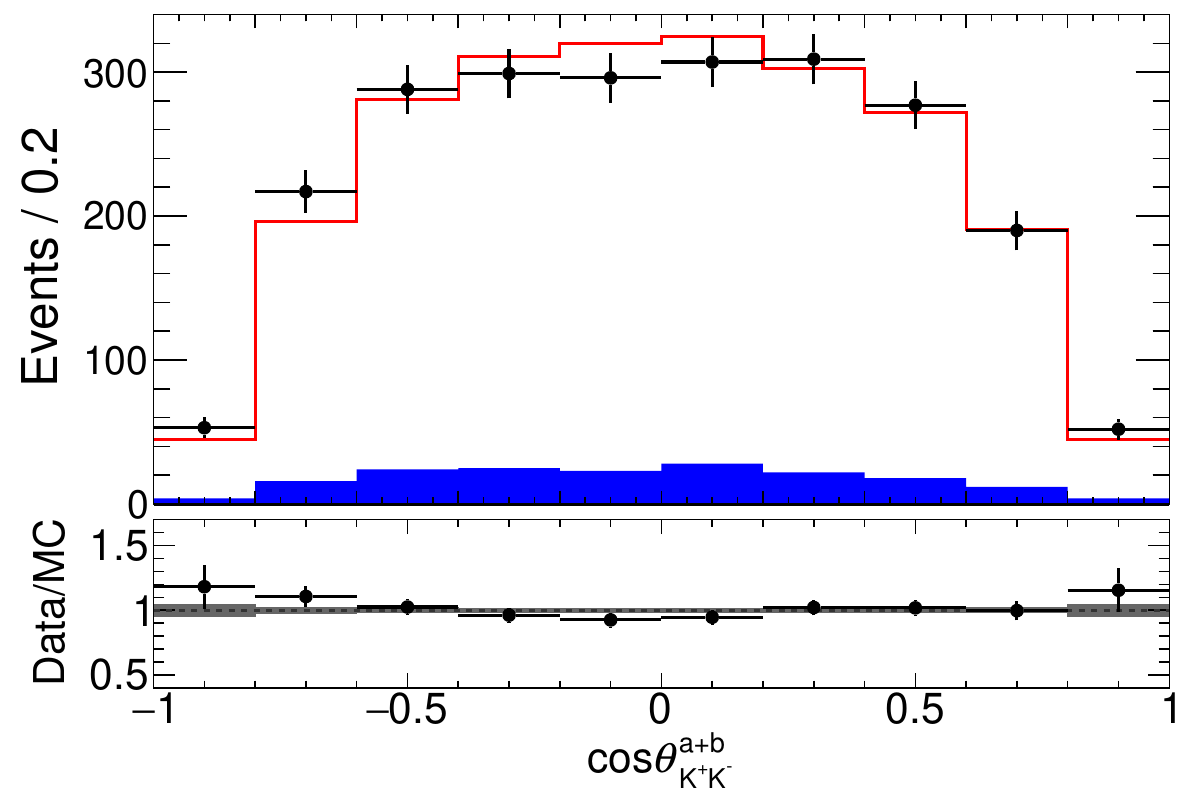}
  \includegraphics[keepaspectratio=true,width=0.495\textwidth,angle=0]{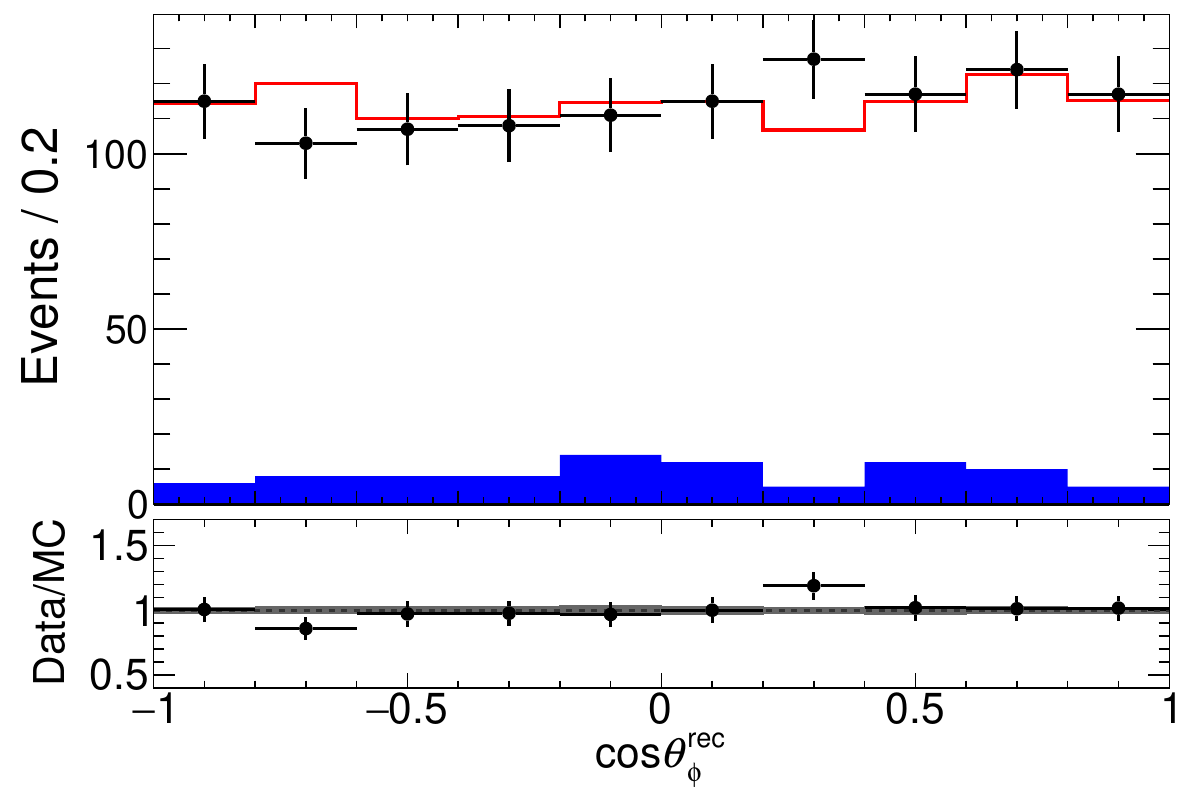}
  \includegraphics[keepaspectratio=true,width=0.495\textwidth,angle=0]{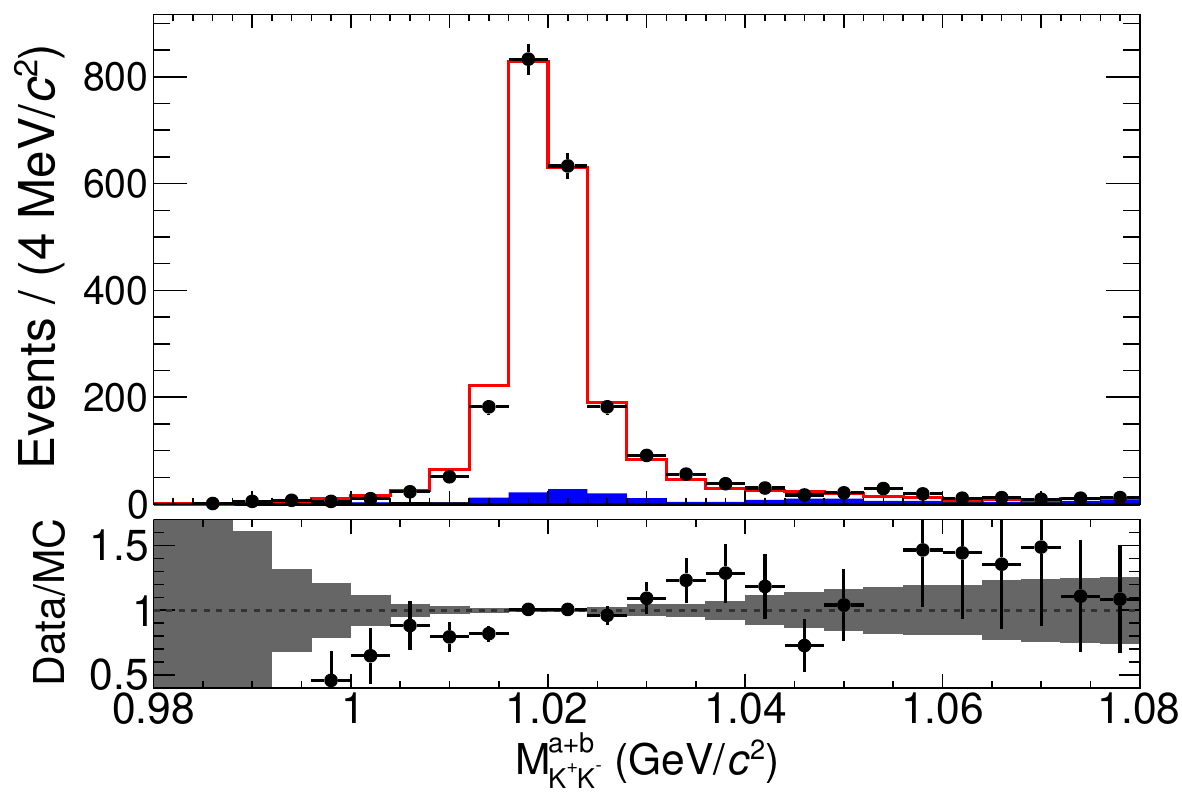}
  \includegraphics[keepaspectratio=true,width=0.495\textwidth,angle=0]{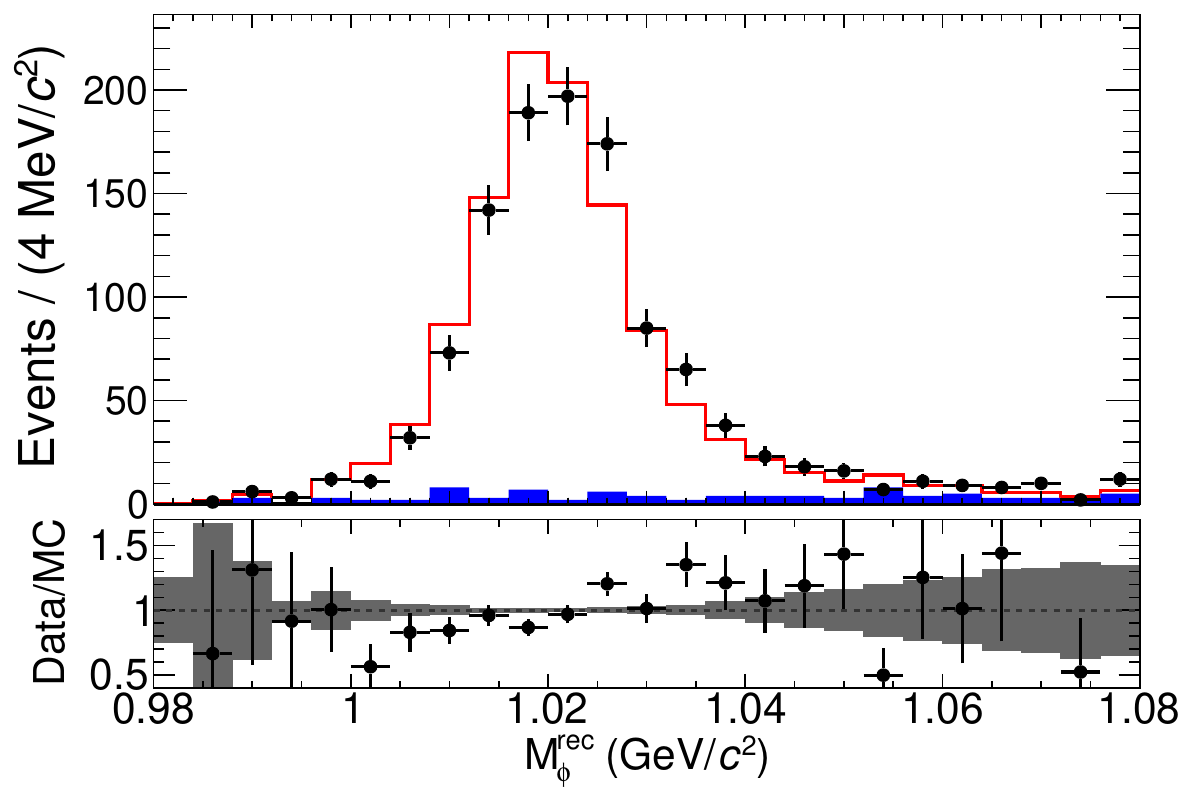}
\caption{Distributions of the momentum, cosines of polar angle and mass spectra of the $\phi$ candidates for partial candidates, two reconstructed $\phi$ candidates are filled in the same histogram (first column). The black points with error bars are data, the solid red lines show the signal MC simulation which is scaled to the total number of events of data, and the blue solid-filled histograms are the background contribution of the inclusice MC sample. The bottom panels show the data and MC comparison, where the error bands indicate the MC statistical uncertainty only.}
\label{fig:compare5}
\end{figure*}

The detection efficiency of the $\psi(3686)\to 3\phi$ decay is evaluated by analyzing the signal MC samples.
Figure~\ref{fig:dalitz} shows the Dalitz plots of the $\psi(3686)\to 3\phi$ candidates selected in data and signal MC samples.
Figures~\ref{fig:compare6} and \ref{fig:compare5} show the comparisons of the momenta and cosines of polar angles of each $\phi$ candidate, as well as the $K^+K^-$ invariant mass spectra between data and MC simulation. The consistency between data and MC simulation is good.
The detection efficiencies of $\psi(3686)\to 3\phi$ is determined to be $(6.13\pm 0.04)\%$ and $(20.58\pm 0.08)\%$ for full and partial reconstruction samples, respectively. An efficiency correction factor is applied to account for the data-MC deviation arising from tracking and PID efficiencies for $K^\pm$, as listed in Sec.~\ref{sec:sys}.

\subsection{Branching fraction}

\begin{figure*}[htbp]\centering
  \begin{tikzpicture}
    \node [ above right, inner sep=0] (image) at (-8,0) {\includegraphics[keepaspectratio=true,width=0.325\textwidth,angle=0]{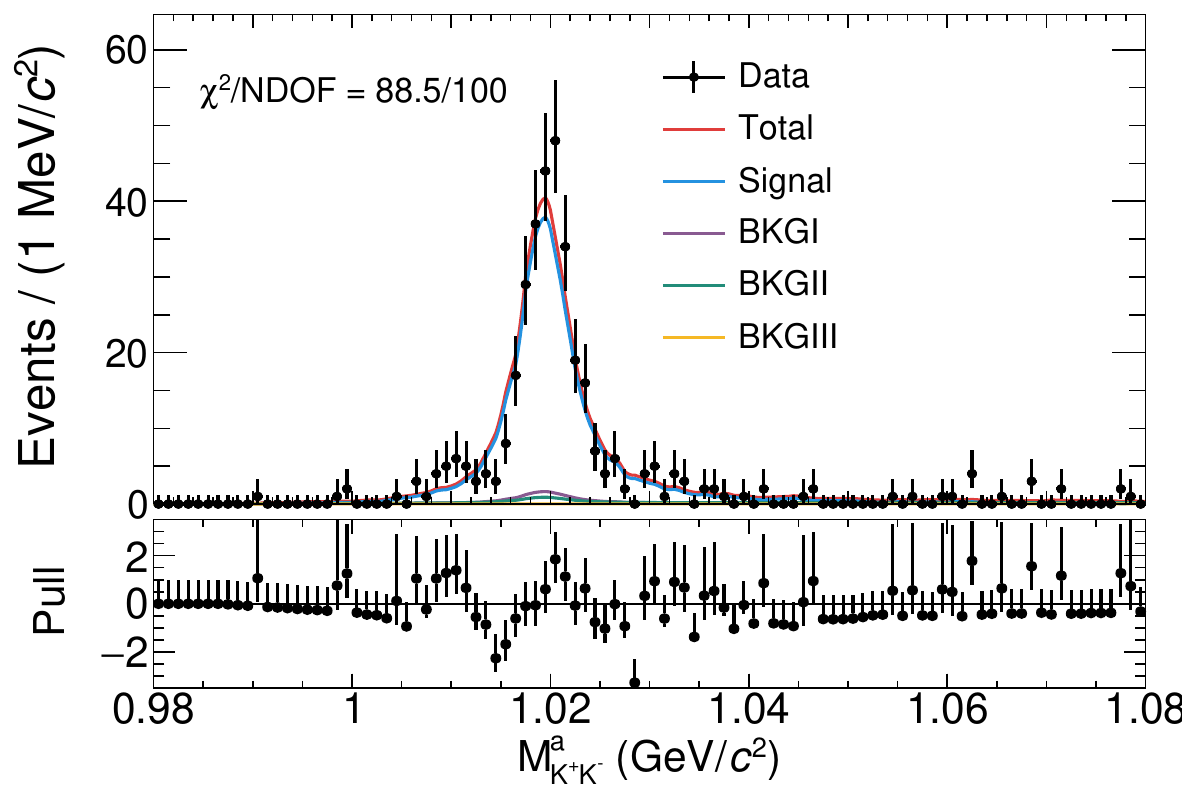}};
    \node [ above right, inner sep=0] (image) at (-2,0) {\includegraphics[keepaspectratio=true,width=0.325\textwidth,angle=0]{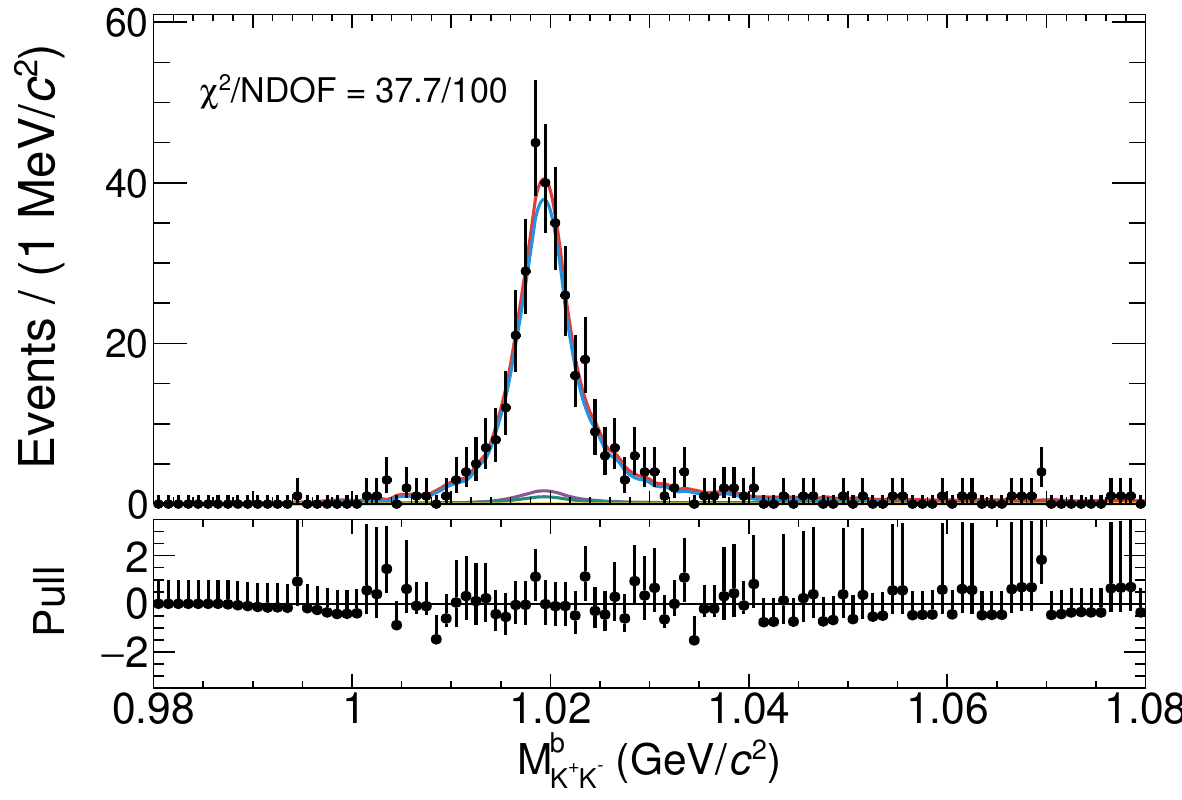}};
    \node [ above right, inner sep=0] (image) at (4,0) {\includegraphics[keepaspectratio=true,width=0.325\textwidth,angle=0]{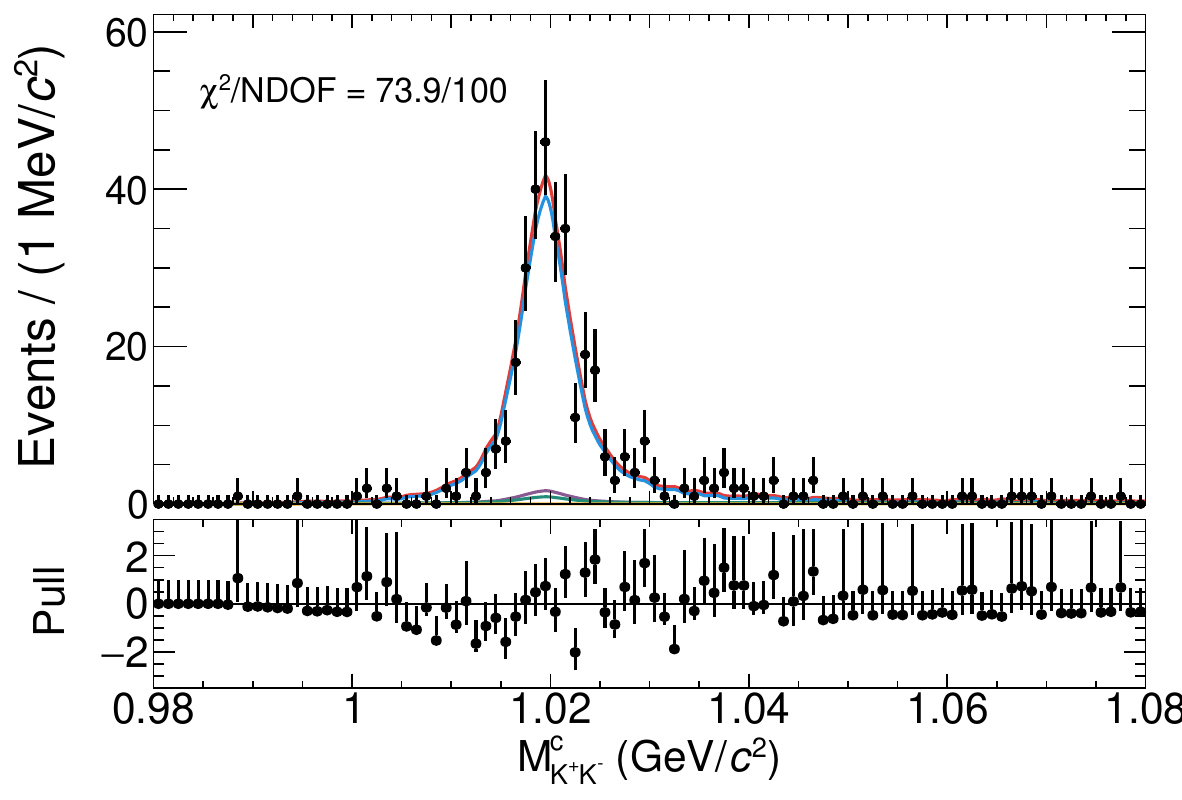}};
    \node [ above right, inner sep=0] (image) at (-8,-4) {\includegraphics[keepaspectratio=true,width=0.325\textwidth,angle=0]{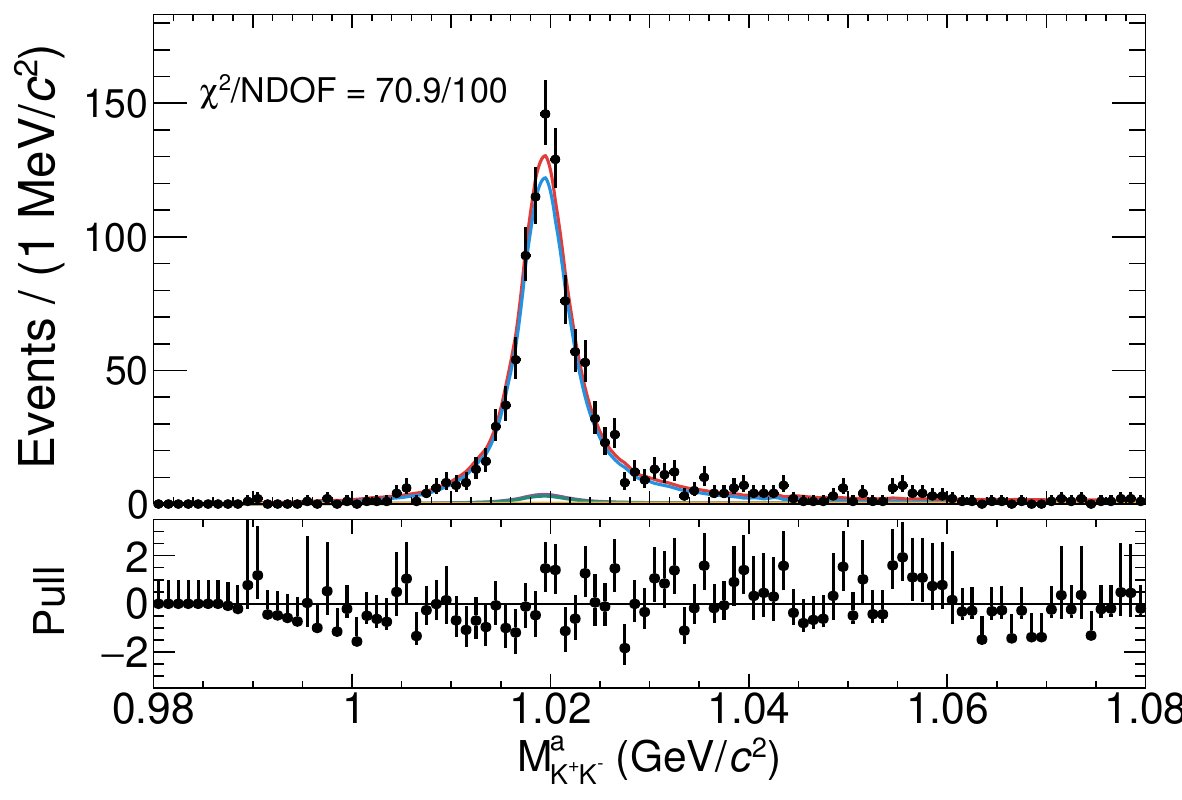}};
    \node [ above right, inner sep=0] (image) at (-2,-4) {\includegraphics[keepaspectratio=true,width=0.325\textwidth,angle=0]{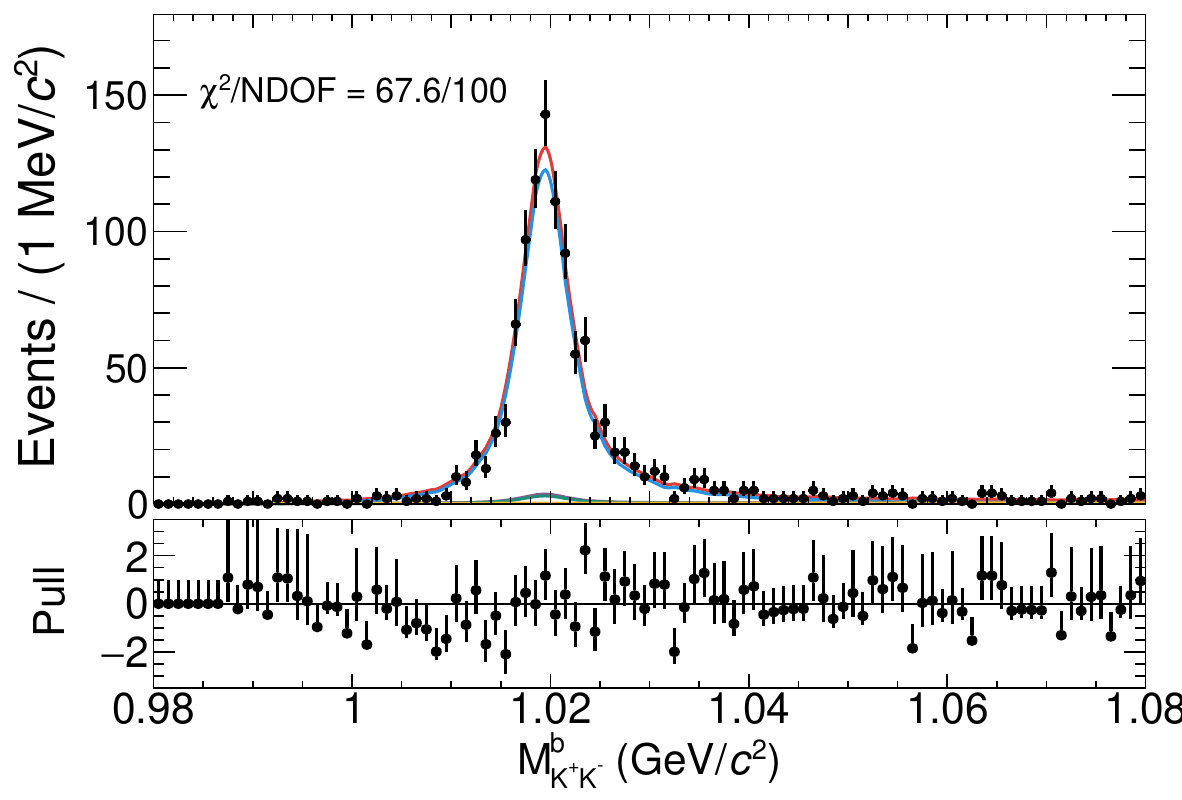}};
    \node [ above right, inner sep=0] (image) at (4,-4) {\includegraphics[keepaspectratio=true,width=0.325\textwidth,angle=0]{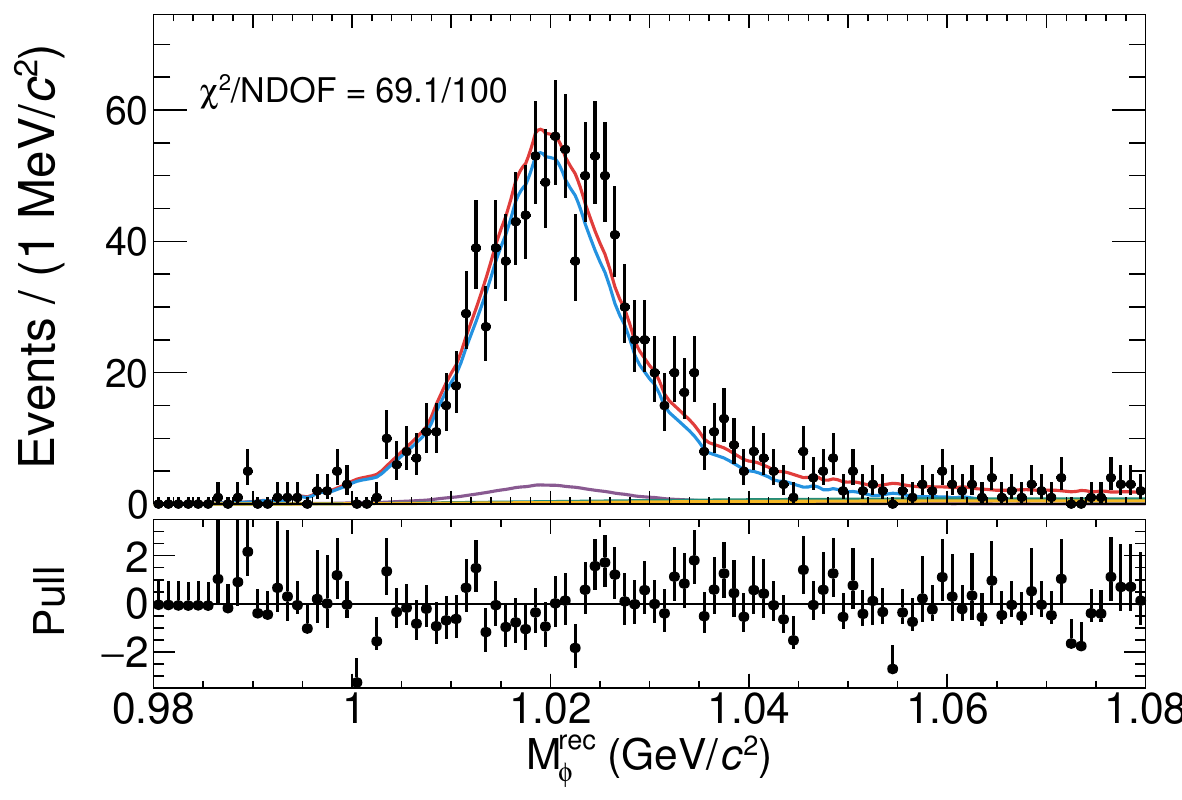}};
  \end{tikzpicture}
  \caption{One-dimensional projections of the simultaneous fit to the $M^a_{K^+K^-}:M^b_{K^+K^-}:M^c_{K^+K^-}(M^{\rm rec}_{\phi})$ distribution of the (top row) full and (bottom row) partial reconstructed candidate events of $\psi(3686)\to 3\phi$ for $\psi(3686)$ data (shown as the dots with error bars). The red solid curves are the total fit results, while the blue curves are the signal contributions of the fit and other curves represent the different background contributions. For each projection, the $\chi^2$/NDOF are provided, with $\chi^2$ being calculated from the difference between the binned data points and the total fit projection, and the NDOF representing the number of bins.}
  \label{fig:simultaneousfit}
\end{figure*}

\begin{figure*}[htbp]\centering
  \begin{tikzpicture}
    \node [ above right, inner sep=0] (image) at (-8,0) {\includegraphics[keepaspectratio=true,width=0.325\textwidth,angle=0]{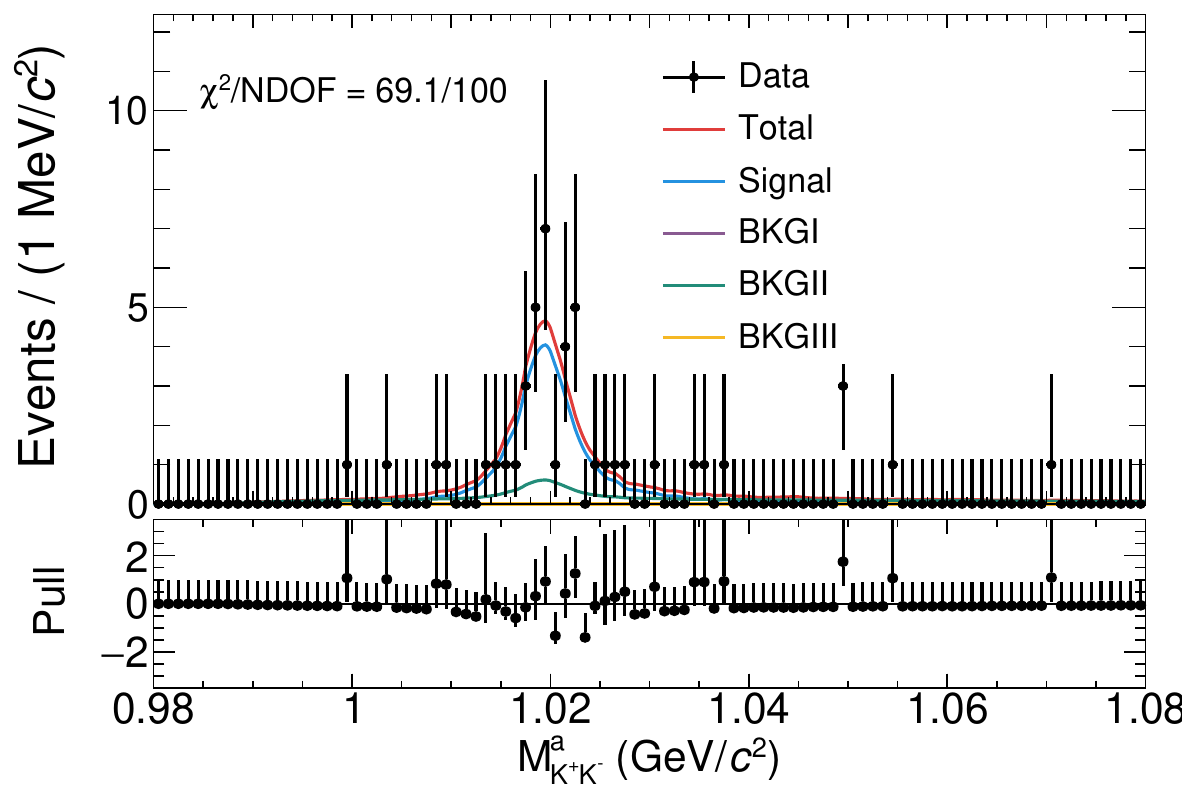}};
    \node [ above right, inner sep=0] (image) at (-2,0) {\includegraphics[keepaspectratio=true,width=0.325\textwidth,angle=0]{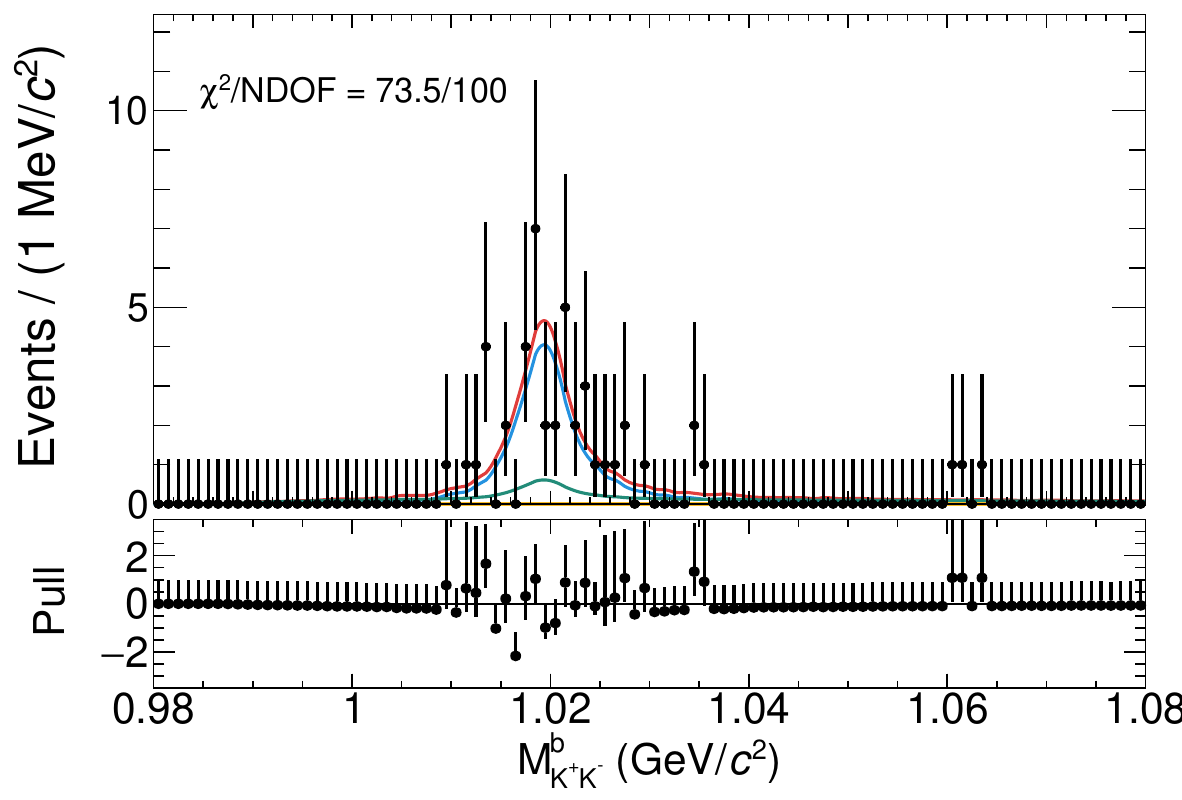}};
    \node [ above right, inner sep=0] (image) at (4,0) {\includegraphics[keepaspectratio=true,width=0.325\textwidth,angle=0]{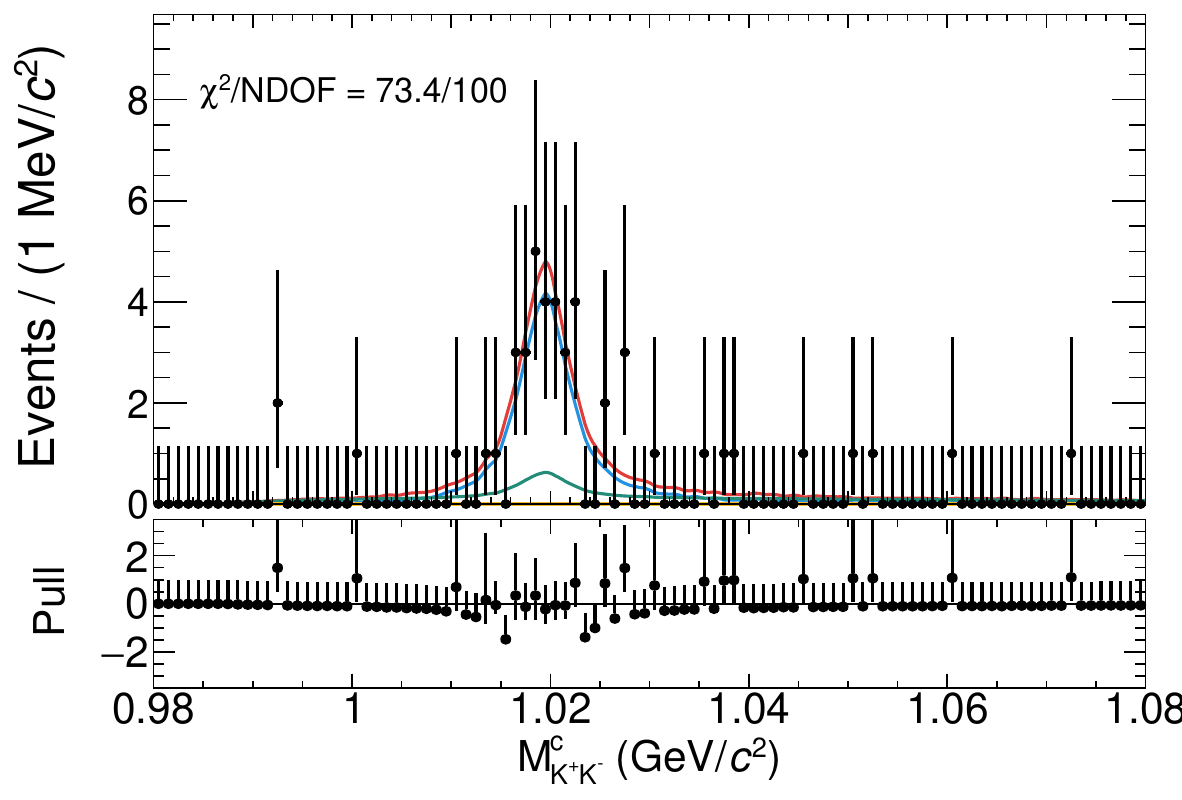}};
    \node [ above right, inner sep=0] (image) at (-8,-4) {\includegraphics[keepaspectratio=true,width=0.325\textwidth,angle=0]{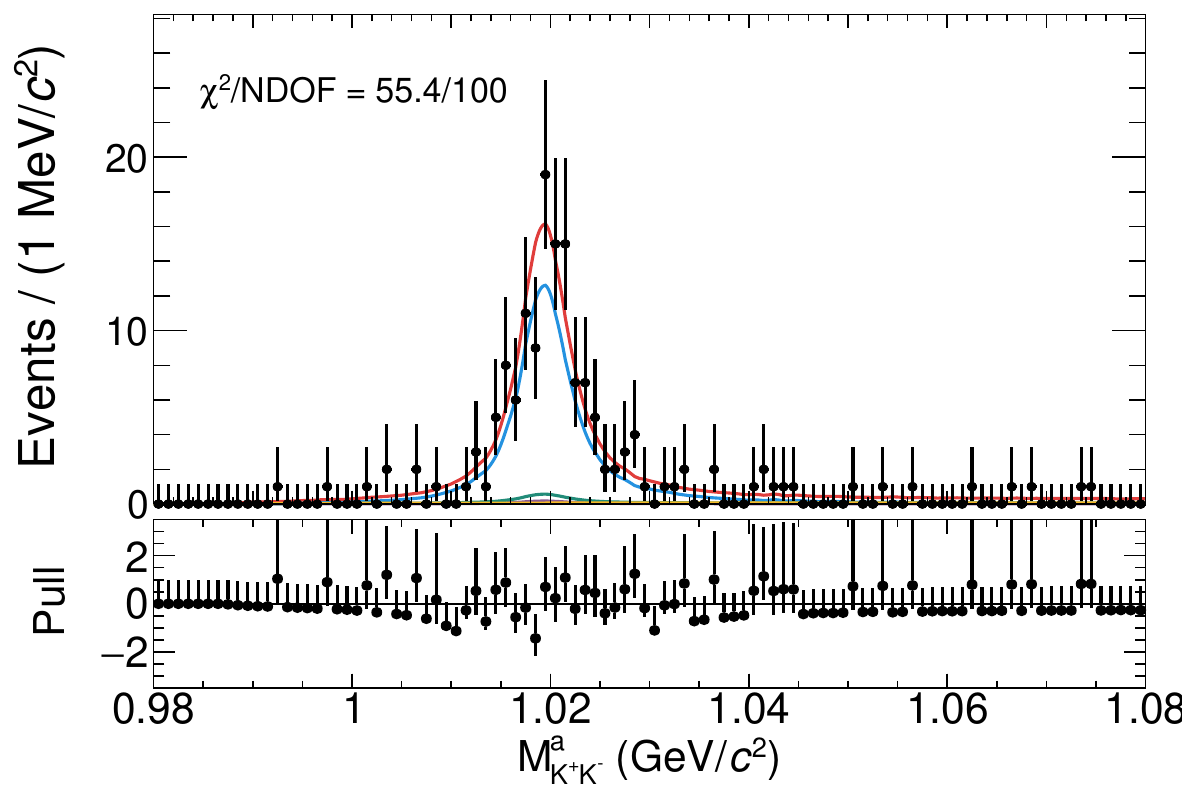}};
    \node [ above right, inner sep=0] (image) at (-2,-4) {\includegraphics[keepaspectratio=true,width=0.325\textwidth,angle=0]{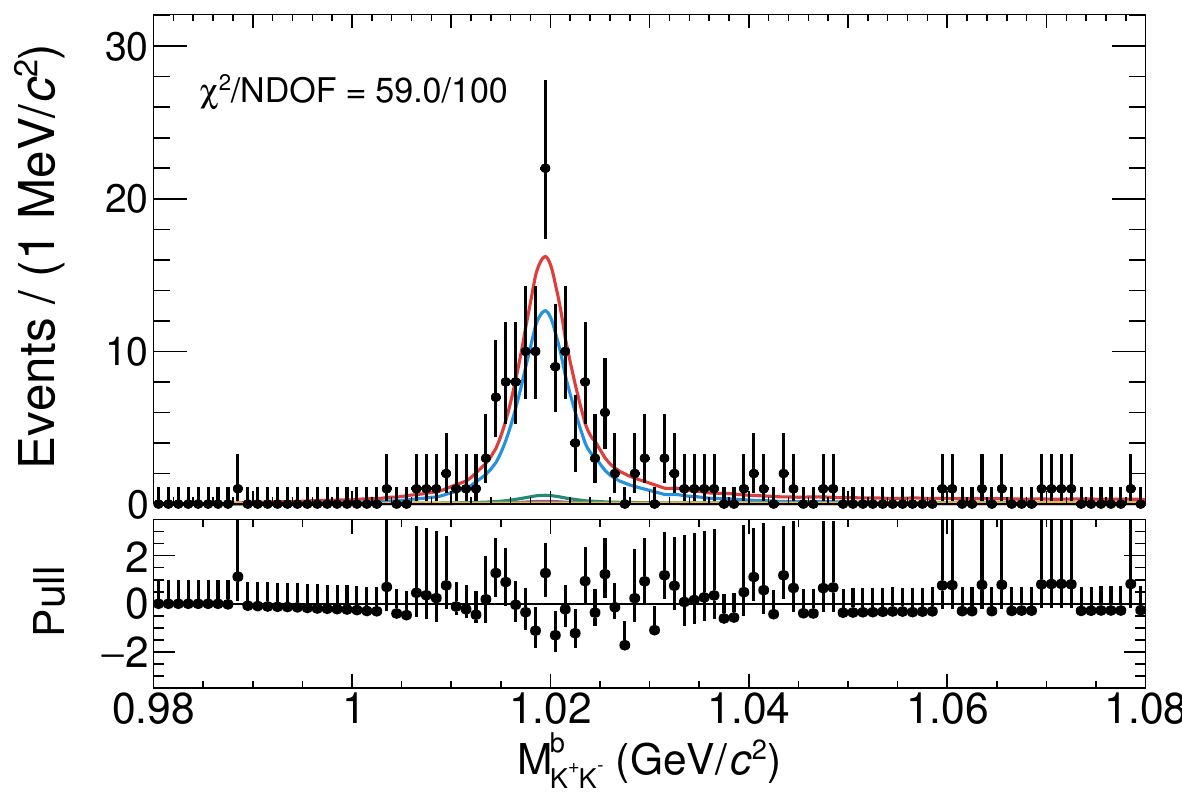}};
    \node [ above right, inner sep=0] (image) at (4,-4) {\includegraphics[keepaspectratio=true,width=0.325\textwidth,angle=0]{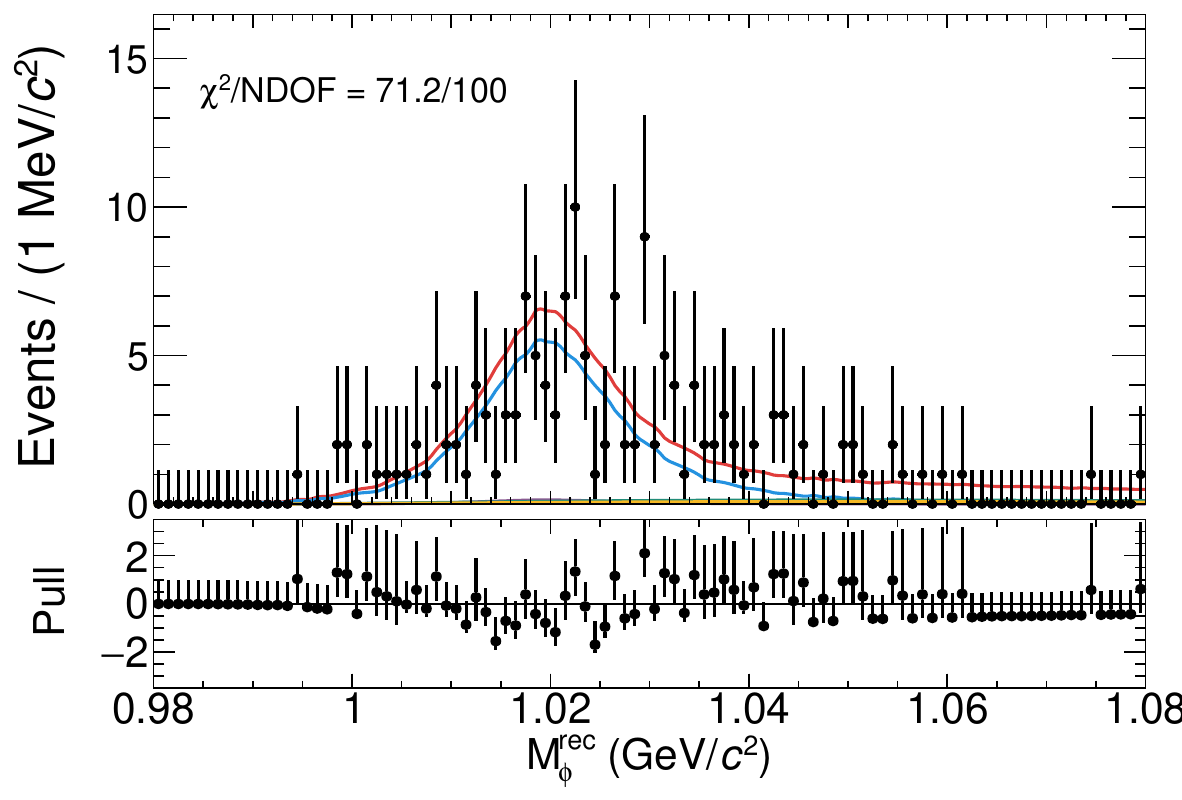}};
  \end{tikzpicture}
  \caption{One-dimensional projections of the simultaneous fit to the $M^a_{K^+K^-}:M^b_{K^+K^-}:M^c_{K^+K^-}(M^{\rm rec}_{\phi})$ distribution of the (top row) full and (bottom row) partial reconstructed candidate events of $e^+e^-\to 3\phi$ for data taken at 3.773 GeV (shown as the dots with error bars). The red solid curves are the total fit results, while the blue curves are the signal contributions of the fit and other curves represent the different background contributions. For each projection, the $\chi^2$/NDOF are given, with $\chi^2$ being calculated from the difference between the binned data points and the total fit projection, and the NDOF representing the number of bins.}
  \label{fig:simultaneousfit3773}
\end{figure*}

Under the assumption that there is no interference between the $\psi(3686)$ and continuum amplitudes,
the branching fraction of the $\psi(3686)\to 3\phi$ decay is determined as follows:
\begin{equation}
    \mathcal{B}_{\psi(3686)\to 3\phi} = \frac{N_{\psi(3686)\to 3\phi} - f_c \times N_{e^+e^-(3773)\to 3\phi}}{N_{\psi(3686)}\mathcal{B}^3_{\phi \to K^+ K^-} \epsilon_{\psi(3686)\to 3\phi}},
\end{equation}
\noindent where $N_{\psi(3686)\to 3\phi}$ and $N_{e^+e^-(3773)\to 3\phi}$ are the numbers of $\psi(3686)\to 3\phi$ and $e^+e^-(3773)\to 3\phi$ events extracted from the data sample taken at $\sqrt{s}=$ 3.686 GeV and 3.773 GeV. They are determined to be $1319\pm 43$ and $138\pm 15$, also the fractions of full reconstructed events out of total events are 23\% and 27\%. Figures~\ref{fig:simultaneousfit} and \ref{fig:simultaneousfit3773} show the fit results on the selected candidates. Meanwhile the fitted event numbers of BKGI, BKGII and BKGIII are $74\pm 24$, $30\pm 16$ and $48\pm 16$.
The factor $f_c$ is introduced to propagate the number of continuum events observed at 3.773 GeV to the 3.686 GeV energy point taking into account luminosities~\cite{psip-num-inc} ${\mathcal L}_{3.686(3.773)}$ and cross sections at different energies.
It is calculated as $f_c =\frac{\epsilon_{\psi(3686)\to 3\phi}}{\epsilon_{e^+e^-(3773)\to 3\phi}}\times\frac{{\mathcal L}_{3.686}}{{\mathcal L}_{3.773}}\times\frac{(3.773~\text{GeV})^{2n}}{(3.686~\text{GeV})^{2n}} $, where $n$ is the power of the $\frac{1}{s}$ dependence of the cross section. We take $n=1$~\cite{dingxiaoxuan} and obtain $f_c = 0.49$.
The total number of $\psi(3686)$ events is labeled as $N_{\psi(3686)}$ and $\mathcal{B}_{\phi \to K^+ K^-}$ is the world average value of the branching fraction of $\phi \to K^+ K^-$ taken from the PDG~\cite{pdg2022}.
The detection efficiencies for $\psi(3686)\to 3\phi$ and the continunum process $e^+e^-(3773)\to 3\phi$ are labeled as $\epsilon_{\psi(3686)\to 3\phi}$ and $\epsilon_{e^+e^-(3773)\to 3\phi}$. The branching fraction of the $\psi(3686) \to 3\phi$ decay is determined to be $(1.46\pm0.05)\times 10^{-5}$, where the uncertainty is statistical only.

The statistical significance is estimated by examining the probability
of the change in negative log-likelihood values when
the signal is included or excluded in the fits. This probability is calculated under the $\chi^2$ distribution hypothesis
taking into account the change in the number of
degrees of freedom. Consequently, the significance is determined to be greater than 10$\sigma$.

We have also examined the Dalitz plot and $\phi\phi$ invariant mass spectra, as shown in Fig.~\ref{fig:dalitz}, and no obvious structure is found.

\section{SYSTEMATIC UNCERTAINTIES}\label{sec:sys}

The systematic uncertainties in the measurement of the branching fraction of the $\psi(3686)\to 3\phi$ are discussed below.

The total number of $\psi(3686)$ events has been determined to be $N_{\psi(3686)}=(2.712\pm0.014)\times 10^9$ with inclusive hadronic events as described in Ref.~\cite{psip-num-inc}. This measurement contributes 0.5\% to the systematic uncertainty of the branching fraction.

The systematic uncertainties of $K^\pm$ tracking and PID are studied with the control sample of $e^+e^-\to \pi^+\pi^- J/\psi ~ (J/\psi \to K^+K^-K^+K^-)$. The differences of $K^\pm$ tracking and PID efficiencies between data and MC simulation are obtained in different transverse momentum intervals.
The data-MC differences are then weighted according to the distribution of the transverse momentum of kaon in the signal decay.
The data to MC ratios of the re-weighted tracking and PID efficiencies are $(98.86\pm0.55)\%$ and $(99.50\pm0.05)\%$, respectively.
Here the errors originate mainly from the limited statistics of the control sample.
The detection efficiency estimated from the MC is corrected with the data to MC ratios, and
the rounded uncertainties of the ratios, $0.6\%$ and $0.1\%$, are taken as the systematic uncertainty of the tracking and PID efficiencies per $K^\pm$, respectively.

The systematic uncertainty of the 3D fit is considered in three aspects.
The background shape is changed from the reversed ARGUS function to a second-order polynominal function.
The signal shape is changed from the simulated MC shape to the shape used by BaBar~\cite{barbarshape},
written as
\begin{equation}
  \sigma(s) = \frac{1}{s^{5/2}}\frac{q^3_{K^+K^-}(s)}{q^3_{K^+K^-}(m_\phi^2)}\left|\frac{\Gamma_\phi m_\phi^3\sqrt{m_\phi\sigma_{\phi\to K^+K^-}/C}}{s - m_\phi^2 + i\sqrt{s}\Gamma_\phi\frac{q^3_{K^+K^-}(s)}{q^3_{K^+K^-}(m_\phi^2)}} \right|^2,
\end{equation}
where $q_{K^+K^-}(s) = \sqrt{s-4m_{K^\pm}^2}$ is a threshold term; $\sigma_{\phi\to K^+K^-}$ is a normalization factor obtained from the fit; $C = 0.389 \times 10^{12}$ nb MeV$^2/c^4$; $m_\phi$ and $\Gamma_\phi$ are the mass and width of the $\phi$ meson.
The alternative fit ranges are chosen as [0.98, 1.09], [0.98, 1.07], [0.97, 1.08], and [0.99, 1.08] GeV/$c^2$.
The quadratic sum of the signal yield variations, 5.7\%, is assigned as the corresponding systematic uncertainty.

The effect of the misidentification of the potential backgrounds from $\psi(3686)\to 2\phi K^+K^-$, $\psi(3686)\to \phi 2(K^+K^-)$ and $\psi(3686)\to 3(K^+K^-)$ to signal are found to be less than 0.1\% and are thereby ignored in the systematic uncertainty.

In the nominal analysis, the helix parameters of charged tracks in the 4C kinematic fit have been corrected with the parameters derived with the control sample of $e^+e^- \to K^*(892) K\pi\to KK\pi\pi$ in Ref.~\cite{helixmethod}. The difference of detection efficiencies with and without helix parameter correction, 1.7\%, is assigned as the corresponding systematic uncertainty.

In the nominal analysis, the $M_{\rm rec}^{K}$ is required to be within
the $\pm 3\sigma$ interval around the kaon mass.
Changing this interval to $\pm 2\sigma$ or $\pm 4\sigma$ results in a relative
change of the measured signal by 1.0\%, which is taken as a systematic uncertainty.

Another source of the systematic uncertainty is the limited MC statistics. This contribution is evaluated as
\begin{equation}
    \centering
    \frac{1}{\sqrt{N}}\sqrt{\frac{(1-\epsilon)}{\epsilon}},
\end{equation}
where $\epsilon$ is the detection efficiency and $N$ is total number of signal MC events. The corresponding number, 0.4\%, is assigned as the systematic uncertainty.

The branching fraction $(49.1\pm0.5)$\% of $\phi \to K^+K^-$ is quoted from the PDG~\cite{pdg2022}, contributing a relative uncertainty of 1.0\%.

In the nominal analysis, we determine the branching fraction without considering the interference between $\psi(3686)$ and continuum amplitudes.
The systematic uncertainty due to this effect is estimated by introducing an  interference term
between psi(3686) and continuum amplitudes. The largest relative change of the signal yield, 9.0\%,
which is observed for $\pm 90^\circ$ phase between the two amplitudes, is taken as the systematic error.

All the systematic uncertainties are summarized in Table~\ref{tab:Sys}. 
The total uncertainty for each reconstruction case in the Table~\ref{tab:Sys} is calculated 
as a quadratic sum of all contributions, which are assumed to be independent within each case.
The total systematic uncertainty is calculated 
using the method described in~\cite{averagemethod}, which takes into account the correlations of systematic 
uncertainties between the different reconstruction cases. 
The uncertainties from the 4C kinematic fit, the $M_{\rm rec}^{K}$ requirement and MC statistics are taken
as uncorrelated and all other contributions are assumed to be fully correlated between the two reconstruction cases.
The total systematic uncertainty is determined to be 11.5\%. 

\begin{table}[htbp]
\centering
    \caption{Systematic uncertainties in the branching fraction measurement.}
    \begin{tabular}{L{3cm} C{2.3cm} C{2.3cm}}
        \hline
        \hline
        Source                            & full~(\%)  & partial~(\%) \\
        \hline
        $N_{\psi(3686)}$                  & 0.5        & 0.5      \\
        Tracking                          & 3.6        & 3.0      \\
        PID                               & 0.6        & 0.5      \\
        3D fit                            & 5.7        & 5.7      \\
        4C kinematic fit                  & 1.7        & none     \\
        $M_{\rm rec}^{K}$ requirement     & none       & 1.0      \\
        MC statistics                     & 0.4        & 0.4      \\
        $\mathcal{B}(\phi \to K^+K^-)$    & 3.0        & 3.0      \\
        Interference                      & 9.0        & 9.0      \\
        \hline
        Total                             & 11.8       & 11.5     \\
        \hline     \hline
    \end{tabular}
    \label{tab:Sys}
\end{table}

\section{SUMMARY}\label{sec:summary}

By analyzing $(2.712\pm0.014)\times 10^9$ events collected with the BESIII detector operating at the BEPCII collider, we report the first observation of the $\psi(3686)\to 3\phi$ decay. The branching fraction of this decay is determined to be $(1.46\pm0.05\pm0.17)\times10^{-5}$, with the first uncertainty being statistical and the second one systematic. Furthermore, we have examined the Dalitz plots and found no obvious structure. Further studies with high statistics data taken at the future super tau-charm factory~\cite{stcf} will be valuable to deeply understand the decay mechanisms of these types of decays and to seek potential new structures in the $\phi\phi$ mass spectrum.

\acknowledgments
The BESIII Collaboration thanks the staff of BEPCII and the IHEP computing center for their strong support. This work is supported in part by National Key R\&D Program of China under Contracts Nos. 2020YFA0406300, 2020YFA0406400; National Natural Science Foundation of China (NSFC) under Contracts Nos. 11635010, 11735014, 11835012, 11935015, 11935016, 11935018, 11961141012, 12025502, 12035009, 12035013, 12061131003, 12192260, 12192261, 12192262, 12192263, 12192264, 12192265, 12221005, 12225509, 12235017; the Chinese Academy of Sciences (CAS) Large-Scale Scientific Facility Program; the CAS Center for Excellence in Particle Physics (CCEPP); Joint Large-Scale Scientific Facility Funds of the NSFC and CAS under Contract No. U1832207; CAS Key Research Program of Frontier Sciences under Contracts Nos. QYZDJ-SSW-SLH003, QYZDJ-SSW-SLH040; 100 Talents Program of CAS; The Institute of Nuclear and Particle Physics (INPAC) and Shanghai Key Laboratory for Particle Physics and Cosmology; European Union's Horizon 2020 research and innovation programme under Marie Sklodowska-Curie grant agreement under Contract No. 894790; German Research Foundation DFG under Contracts Nos. 455635585, Collaborative Research Center CRC 1044, FOR5327, GRK 2149; Istituto Nazionale di Fisica Nucleare, Italy; Ministry of Development of Turkey under Contract No. DPT2006K-120470; National Research Foundation of Korea under Contract No. NRF-2022R1A2C1092335; National Science and Technology fund of Mongolia; National Science Research and Innovation Fund (NSRF) via the Program Management Unit for Human Resources \& Institutional Development, Research and Innovation of Thailand under Contract No. B16F640076; Polish National Science Centre under Contract No. 2019/35/O/ST2/02907; The Swedish Research Council; U. S. Department of Energy under Contract No. DE-FG02-05ER41374.

\end{document}